\documentclass[aps,twocolumn,showpacs,groupedaddress]{revtex4-1}
\usepackage{amssymb,amsmath}
\usepackage{graphicx,array}
\usepackage{dcolumn}
\usepackage{bm}
\usepackage{float}

\begin{document}

\tableofcontents
\clearpage

\title{The Role of Density Functional Theory Methods in the Prediction 
              of Nanostructured Gas-Adsorbent Materials}

\author{Claudio Cazorla}
\email{c.cazorla@unsw.edu.au}
\affiliation{School of Materials Science and Engineering, University of New South Wales, Sydney NSW 2052, Australia \\
Integrated Materials Design Centre, University of New South Wales, Sydney NSW 2052, Australia}

\begin{abstract}
With the advent of new synthesis and large-scale production technologies,
nanostructured gas-adsorbent materials (GAM) such as carbon nanocomposites
and metal-organic frameworks are becoming increasingly more influential in our 
everyday lives. First-principles methods based on density functional theory (DFT) 
have been pivotal in establishing the rational design of GAM, a factor which 
has tremendously boosted their development. 
However, DFT methods are not perfect and due to the stringent accuracy 
thresholds demanded in modeling of GAM (e.g., exact binding energies to within 
$\sim 0.01$~eV) these techniques may provide erroneous conclusions in some 
challenging situations. Examples of problematic circumstances include 
gas-adsorption processes in which both electronic long-range exchange and nonlocal 
correlations are important, and systems where many-body 
energy and Coulomb screening effects cannot be disregarded. In this critical 
review, we analyze recent efforts done in the assessment of the performance of 
DFT methods in the prediction and understanding of GAM. Our inquiry is constrained 
to the areas of hydrogen storage and carbon capture and sequestration, for which 
we expose a number of unresolved modeling controversies and define a set of 
best practice principles. Also, we identify the subtle problems found 
in the generalization of DFT benchmark conclusions obtained in model cluster systems 
to real extended materials, and discuss effective approaches to circumvent them. The 
increasing awareness of the strengths and imperfections of DFT methods in the 
simulation of gas-adsorption phenomena should lead in the medium term to more 
precise, and hence even more fruitful, \emph{ab initio} engineering of GAM.
\end{abstract}


\maketitle

\section{Introduction}
\label{sec:intro}

\begin{figure*}
\centerline{
\includegraphics[width=1.00\linewidth]{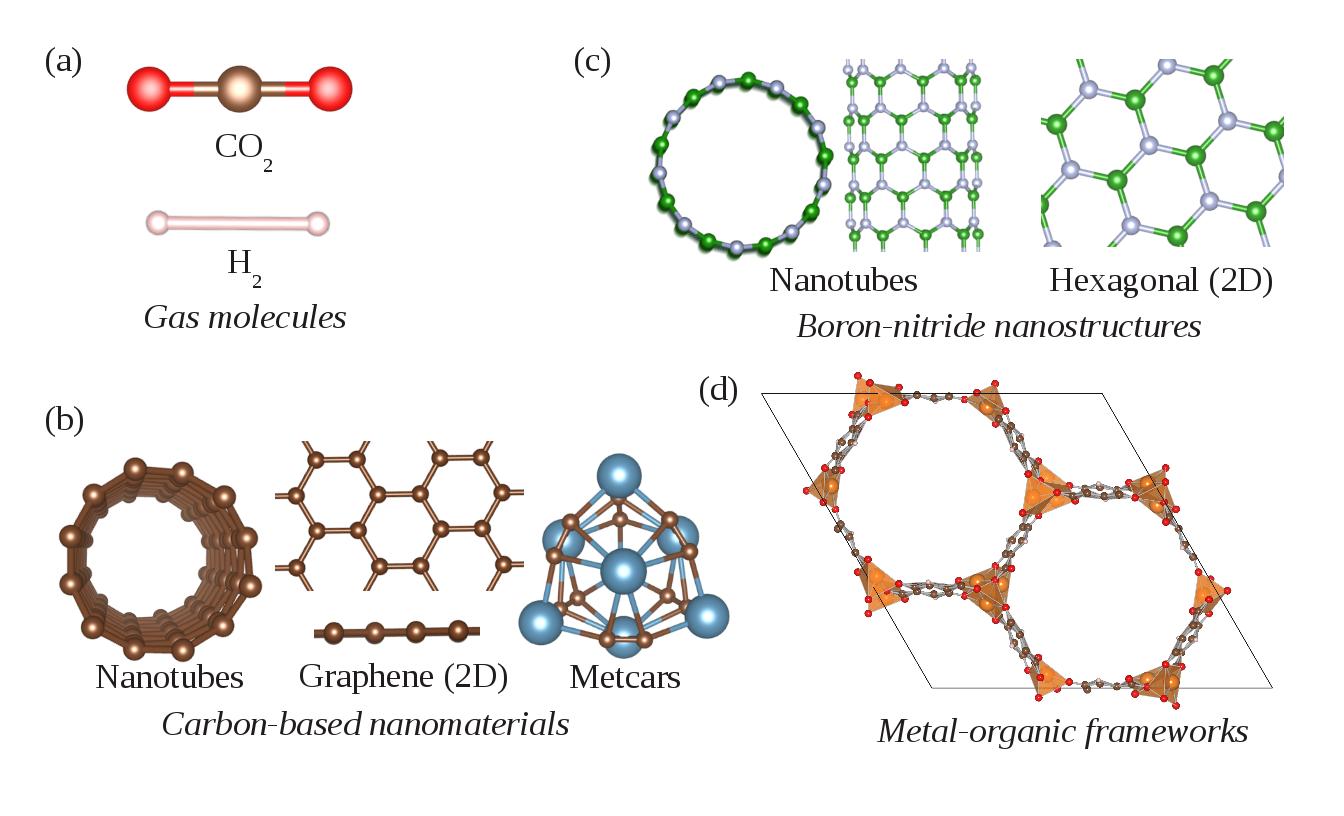}}
\vspace{-0.75cm}
\caption{(a) Representation of the gas species considered in this review and (b)-(d) 
        some of the most popular families of nanostructured gas-adsorbent materials: carbon-based 
        nanomaterials, boron-nitride nanostructures and MOF (see text).}   
\label{figintro-1}
\end{figure*}

\subsection{Rational design of gas-adsorbent materials}
\label{subsec:design}
Nanostructured gas-adsorbent materials (GAM) are the cornerstones of potentially 
revolutionary advancements in critical and fast growing technological fields 
like molecular sensing, energy storage and harvesting, and environmental and sustainability 
engineering. Their exceptional high surface to volume ratio, regular atomic composition, 
tunable reactivity, transport properties, and assembling affinity to form supramolecular  
systems, have permitted the realization of timely and cost-effective applications like,
for instance, the detection and removal of toxic substances from water and air, dense storage 
of hydrogen and natural gas in solid state matrices, sequestration of carbon dioxide from flue 
gases generated in electricity production plants, design of high-performance photovoltaic cells, 
and enhanced long-lasting operation of batteries, to cite just a few examples~\cite{dai12,bonaccorso10,yavari12,zhao12,sahoo12}.

Popular families of nanostructured GAM include zeolites, metal oxides nanocrystals (e.g., CaO and 
Al$_{2}$O$_{3}$), carbon-based nanomaterials (CN, e.g., nanotubes, sheets, met-cars, graphite 
intercalation compounds, and frameworks of organic pillared graphene), metal hydrides nanoparticles 
(e.g., MgH$_{2}$ and LiBH$_{4}$), and covalent- and metal-organic frameworks (COF and MOF). Of all 
these species CN and MOF (see Fig.~\ref{figintro-1}) stand out as some of the most promising 
GAM for energy and environmmetal applications, particularly to what concerns the capture and 
storage of hydrogen (H$_{2}$), carbon dioxide (CO$_{2}$), and methane 
gases~\cite{mendoza12,fracaroli14,furukawa09,broom13,urbonaite08,tranchemontagne12,yurum09,dietzel09,kumar14,simmons11}.

Currently reported GAM gas-selectivity and storage capacities, however, still 
remain below the stringent commercial targets set by specialized government bodies
and agencies. For instance, hydrogen storage systems need to achieve an overall 
capacity of $5.5$~wt\% hydrogen with a volumetric ratio of $40$~g/L for competitive 
vehicle applications~\cite{us11}, and capture of the $90$~\% of the carbon dioxide produced 
in the generation of electricity must be reached within less than a $35$~\% of increase in 
the final costs~\cite{rackley10,hester10}. The search for optimal gas-adsorption processes
and GAM, therefore, remains an area of very active scientific and technological 
research.  

A key aspect for potential GAM to be successful is to find the optimal chemical 
compositions and pore topologies to work under specific thermodynamic conditions. 
The number of possible stoichiometric and structural GAM configurations is tremendously 
large, hence in practice systematic experimental searches based on trial-error 
strategies turn out to be cumbersome and very inefficient. Rational engineering of gas-adsorbent 
interactions at the atomic scale, represents a key notion to achieve success on such a design
grand challenge in the short and middle term. In this context, computational simulation methods
emerge as invaluable theoretical tools for the screening and rational engineering of auspicious 
GAM.

\subsection{Computational simulation techniques for modeling of GAM}
\label{subsec:simulation}
Common simulation techniques in the study of GAM can be classified into two 
major categories: ``semi-empirical'' and ``first-principles''.
In semi-empirical approaches, the interactions between atoms 
are modeled with analytical functions, known as force fields or classical 
potentials, which are devised to reproduce a certain 
amount of experimental data or the results of high-accuracy 
calculations. The inherent simplicity of classical potentials 
makes it possible to address the study of GAM and gas-adsorption
processes considering realistic thermodynamic conditions and length/time
scales, with well-established simulation techniques like, for instance, 
molecular dynamics and grand canonical Monte Carlo~\cite{frenkel}. 
With the current computational power and algorithm development, key 
features in GAM which are directly comparable to observations (e.g., 
adsorption isotherms, elastic properties and diffusion 
coefficients~\cite{skoulidas04,xiang10,battisti11,garberoglio12,garberoglio15}) 
can be computed routinely on a standard office computer. 
Also, the relevance of quantum nuclear effects in gas-adsorption phenomena can 
be estimated accurately with semi-empirical approaches~\cite{gordillo11b,cazorla13b,carbonell13,gordillo11}. 
Nevertheless, in spite of the great versatility of semi-empirical methods, classical 
potentials may present impeding transferability issues in certain situations. 
This type of drawbacks is related to the impossibility of mimicking the features 
of the targeted material at conditions different from those in which the setup 
of the corresponding force field was performed~\cite{csanyi04,dzubak12}.   

In this context, the outputs of first-principles calculations turn out to be crucial. 
As the name indicates, empirical information is not contained on first-principles methods,
also known as \emph{ab initio}. The interactions between atoms are directly obtained 
from applying the principles of quantum mechanics to the electrons and nuclei. 
Transferability issues, therefore, are totally missing in \emph{ab initio} approaches.
Examples of first-principles techniques include, density functional theory 
and quantum Monte Carlo, to cite just a few. Although these approaches are very accurate, 
they can be also very demanding in terms of computational expense. This circumstance poses
serious difficulties to the study of kinetic and thermodynamic effects in extended GAM 
with \emph{ab initio} techniques. Common acceleration schemes within first-principles methods 
entail the use of pseudopotentials~\cite{vanderbilt90,troullier91}. 
Many materials properties can be predicted basing exclusively on the behaviour of 
the valence electrons, and by employing pseudopotential techniques explicit treatment 
of the core electrons is avoided in the simulations. Pseudopotentials can actually be the source 
of potential errors, however they make also the simulation of heavy atoms and medium-size systems 
feasible. Fortunately, some strategies can be used to minimize the impact of the approximations 
introduced by pseudopentials like, for instance, the projector augmented wave~\cite{blochl94} and 
linearized augmented plane wave methods~\cite{andersen75}. 
In the present critical review, we will concentrate on analyzing specific aspects of the simulation 
of GAM with first-principles methods, obviating the outcomes of indispensable and physically 
insightful semi-empirical approaches.

\subsection{Capabilities of density functional theory methods}
\label{subsec:dft}
Standard density functional theory (DFT) techniques (i.e., based on local and semilocal 
approximations to the electronic exchange-correlation energy, typically LDA and GGA) have 
become the \emph{ab initio} methods of choice in the study and design of nanostructured GAM. 
These techniques have been demonstrated to reproduce with notable accuracy the 
interplay of a wide range of interactions in hundreds-of-atoms systems, while keeping within 
reasonably affordable limits the accompanying computational expense. However, standard DFT 
methods present some well-known limitations in describing gas-adsorption phenomena occurring 
in low-coordinated atomic environments (e.g., the binding of small molecules to surfaces 
and cavities). 
For instance, due to the local nature of the employed energy functionals standard DFT 
methods cannot reproduce the electronic long-range correlations resulting from 
instantaneously induced dipole-dipole, dipole-quadrupole, quadrupole-quadrupole and so 
on interactions. Another common DFT fault is the presence of electronic self-interaction errors, 
which derive from an imperfect cancellation between the auto-correlation (i.e., spurious interaction 
of an electron with itself) and exchange energies. This type of errors can alter dramatically the 
description of charge-transfer complexes, chemical reactions, and electron affinities~\cite{perdew81,lee88,patchkovskii02,cohen11}. 
Fortunately, many developments have been realized during the last two decades which have solved 
part of these drawbacks (e.g., nonlocal and hybrid exchange-correlation energy functionals, 
see next section), permitting so to achieve remarkable agreement between theory and experiments. 
Nevertheless, a number of critical aspects remain yet challenging to customary DFT methods 
that could be hindering the rational design of GAM. 

\subsection{DFT challenges in GAM design}
\label{subsec:dftchallenges}
\begin{figure}
\centerline{
\includegraphics[width=1.00\linewidth]{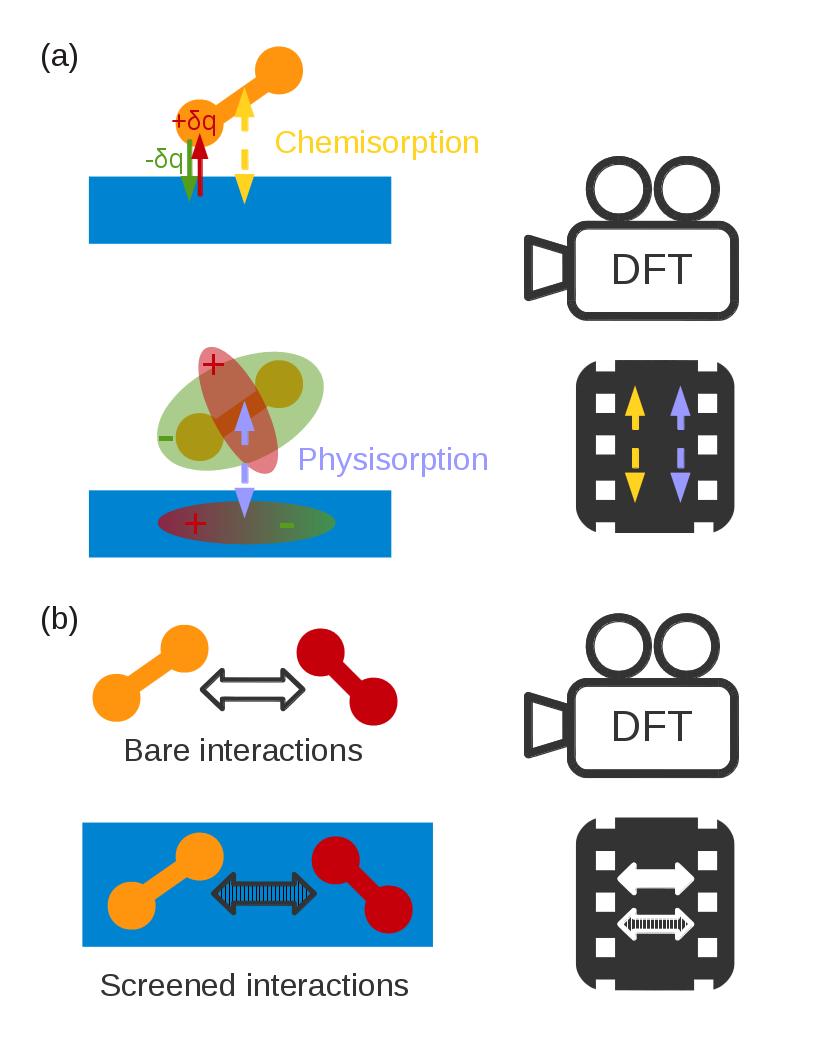}}
\vspace{-0.50cm}
\caption{Schematic representation of the two main challenges of customary DFT methods 
         in GAM design: (a)~reproduction of covalent and dispersion 
         interactions on an equal footing and (b)~description of Coulomb 
         screening and many-body energy interactions in extended systems.}
\label{figintro-2}
\end{figure}

One of such challenges consists in accounting for the long-range electronic correlations 
while simultaneously amending the electronic self-interaction errors (see Fig.~\ref{figintro-2}a). 
Along with the adsorption of gas molecules on surfaces, regions of significant electron depletion 
and accumulation may appear which induce strong electrostatic interactions and the shift of electronic 
energy levels. In those conditions, one could expect that by adding a portion of the exact Hartree-Fock 
exchange energy to the selected DFT functional, in order to lower the electron self-interaction errors 
to negligible levels, accurate binding energies will follow. Reported evidence, however, suggests 
that accounting for weak dispersion interactions in charge-transfer processes may turn out to be   
also decisive~\cite{johnston08,steinmann12b}. Unfortunately the causes behind such an 
unexpected result are not yet totally understood, due in part to the difficulties 
encountered in the decomposition of DFT energies into fundamental portions.
Rationalization of this complexity clearly is needed for optimizing the computational 
load associated to DFT modeling of GAM (i.e., accounting for dispersion interactions 
normally requires intensive calculations), and also for ensuring the reliability of 
prospective and already published simulation works on the storage and sequestration 
of gases. 
      
Another important DFT threat consists in accounting adequately for the screening of  
bare interactions in periodically extended systems (see Fig.~\ref{figintro-2}b). 
Many-body energy and Coulomb screening effects can be equally important in physisorption and 
chemisorption phenomena, and the issues encountered in their description are a consequence 
of the pairwise additivity assumed in the construction of most DFT functionals. 
Essentially, the interaction energy between two atoms remains unaltered no matter 
what medium separates them or what collective excitations happen in the material~\cite{misquitta10,tkatchenko12,gobre13}. 
In this context, the adiabatic connection fluctuation-dissipation (ACFD) theorem has 
been exploited to calculate correlation DFT energies that incorporate many-body 
higher-order terms. This is the case of the random phase approximation to DFT 
(RPA-DFT)~\cite{dobson06,angyan11,hesselmann12} and the DFT+MBD~\cite{ruiz12,tkatchenko12b,ambrosetti14,distasio14} 
methods (see Sec.~\ref{sec:dft}), which at present are receiving the highest attention. 
Nonetheless, the development of many-body DFT-based methods are still on their infancy 
and the associated computational expenses are elevated (typically ranging from two to four 
orders of magnitude larger than standard DFT). The corresponding degree of applicability 
therefore remains still limited.          

\subsection{Inherent complexity of DFT benchmark studies}
\label{subsec:benchmarking}
Most of \emph{ab initio} works published to date on the design of GAM are based 
on standard DFT calculations (see Sec.~\ref{sec:discussion} for 
more details). Standard DFT methods are very effective in dealing with large atomic systems 
and a myriad of user-friendly DFT packages, practically available to everyone nowadays, have 
facilitated their widespread use. However, chemical accuracy is generally demanded on 
GAM design (e.g., correct binding energies to within $\sim 0.01$~eV) and thus, for the reasons 
highlighted in the previous section, employing standard DFT methods may not always be adequate. 

In fact, to carry out DFT benchmark tests on the assessment of GAM is of paramount importance 
for rigorously establishing acceptable balances between computational feasibility and predictive 
reliability. Performing computational studies of such a type, however, is not a trivial task. 
First, one has to be familiar with the technical and foundational aspects of highly accurate 
quantum chemistry methods (QCM) like M\o ller-Plesset perturbation theory (MP2), the coupled-cluster 
method with single, double and perturbative triple excitations [CCSD(T)], and quantum Monte Carlo 
(QMC), to cite just a few. In those approaches the Schr\"{o}dinger equation of the many-electron 
system of interest is solved directly by handling the corresponding wavefunction. Many-electron 
wavefunctions generally are expressed in real space as a linear combination of atomic orbitals (LCAO). 
LCAO do not fulfill the conditions of Bloch functions and thus the simulation of periodic systems 
like crystals and surfaces, although possible, it is not straightforward with them~\cite{manby10,pisani80,dovesi83,causa88}. 
Consequently, most QCM studies are performed in supercells containing model cluster systems 
where periodic boundary conditions are not applied. Meanwhile, QCM are inherently complex and 
the associated computational load in general scales poorly with the number of electrons (namely 
as $N^{4-7}$). 

\begin{figure}
\centerline{
\includegraphics[width=1.00\linewidth]{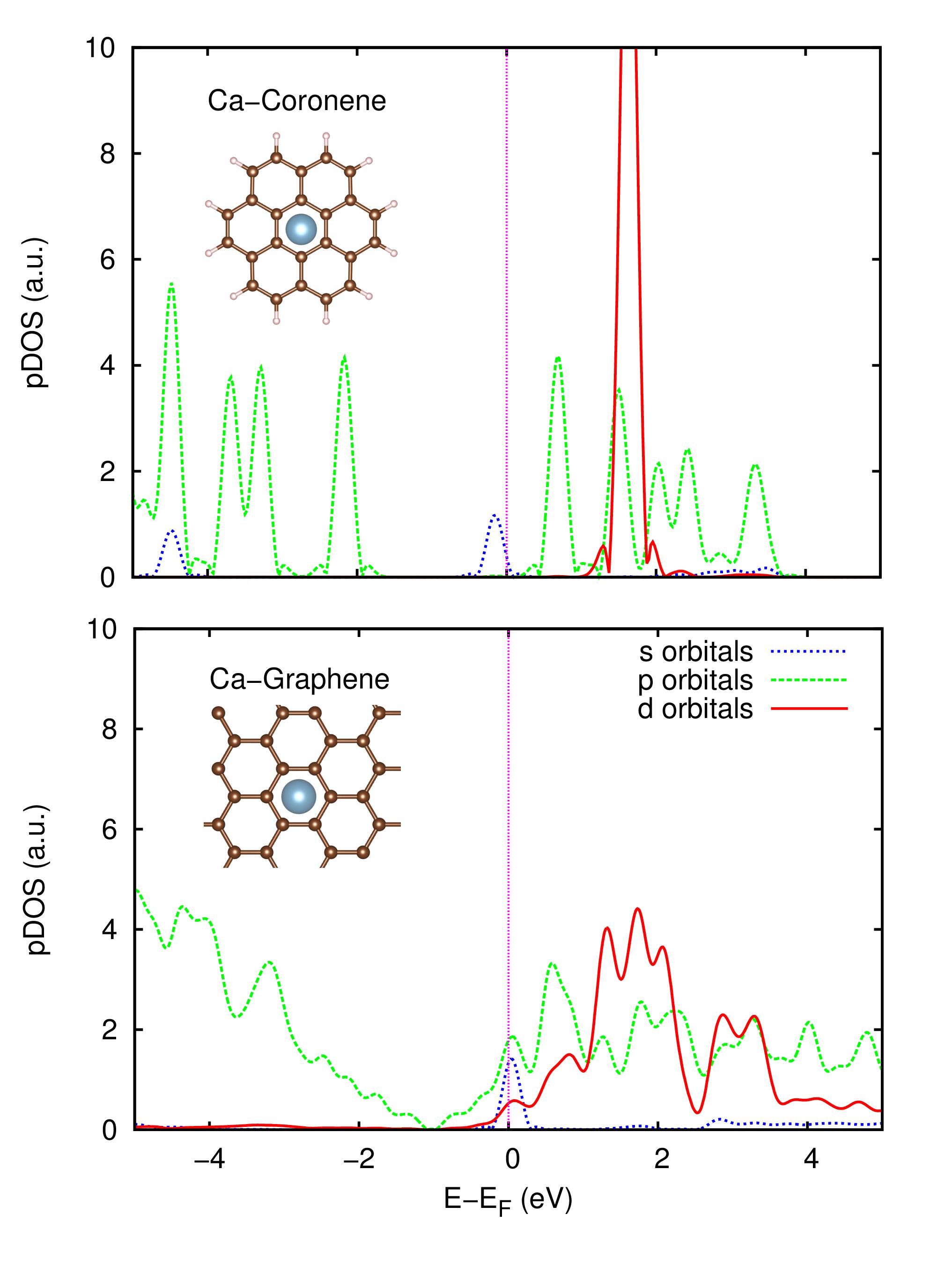}}
\vspace{-0.40cm}
\caption{Partial density of $s$, $p$, and $d$ electronic states calculated in a
         Ca-C$_{24}$H$_{12}$ molecule (\emph{Top}), and in Ca-doped graphene (\emph{Bottom}).
         The ratio of calcium to carbon atoms is the same in both 
         systems, namely $1/24$. Calculations have been performed with standard DFT 
         methods, and the Fermi energy level in both systems has been shifted to zero. 
         (This figure has been adapted from work.~\cite{cazorla09}.)}
\label{figintro-3}
\end{figure}

Besides the technical intricacies, a conceptual problem also arises when performing benchmark studies 
in model cluster systems: on what grounds is correct to generalize the so-obtained conclusions to 
realistic extended materials? 
To help in answering this question, we show in Fig.~\ref{figintro-3} the partial density of 
electronic states (pDOS) calculated in graphene, i.e., an extended one-atom thick carbon surface, 
decorated with Ca impurities, and a coronene molecule doped with a Ca adatom (i.e., Ca@C$_{24}$H$_{12}$) 
using standard DFT methods (in both systems the ratio of Ca impurities to carbon atoms is the same). 
We note that the coronene molecule is generally considered as a large enough system in which 
to carry out computational accuracy tests. However, as it can be appreciated in the figure the 
pDOS computed in coronene differs greatly from the one obtained in the ``equivalent'' 
graphene-based system (e.g., compare the distribution of electronic $d$ states around the Fermi 
energy level in the two cases). Also, charge-transfer calculations based on Bader's theory show 
that the Ca adatom donates about $0.08$~$e^{-}$ to the coronene molecule (i.e., weak ionic interaction), 
whereas in Ca-decorated graphene it supplies $\sim 0.9$~$e^{-}$ to the carbon surface (i.e., 
strong ionic interaction). In fact, quantum confinement effects can introduce important differences 
in the electronic structure of apparently similar extended and cluster systems, thereby leading 
to completely unlike gas-adsorption behaviours in the two situations~\cite{ma11,smith13}. 
In addition to this, long-range electronic exchange and nonlocal correlations, which are ubiquitous 
in gas-adsorption phenomena, normally turn out to be disguised in model cluster systems. 

An obvious solution to overcome these likely scaling problems, affecting most benchmark studies, consists 
in using highly accurate quantum chemistry approaches which, alike to DFT, can handle the simulation of 
periodic systems. All the necessary benchmark and DFT calculations therefore can be undertaken in a same 
extended simulation cell. In this regard, quantum Monte Carlo (QMC) emerges as one of the most promising 
family of methods for benchmarking DFT. Actually, the formalism of many-electron Bloch functions has been
established within QMC since long time ago~\cite{rajagopal95,foulkes01} and several implementations of this 
approach are already available in open-source packages like, for instance, CASINO~\cite{needs10}, 
QWALK~\cite{wagner09}, and QMCPACK~\cite{qmcpack}. 
Also, the balance between accuracy and scalability in QMC is among the highest of all quantum chemistry methods 
(e.g., the accompanying computational cost climbs as $N^{2-3}$ with the number of electrons~\cite{esler08}). 
It is worth mentioning that very recent methodological advances have permitted also the implementation 
of the MP2 and the CCSD approaches in a planewave framework~\cite{marsman09,gruneis10,gruneis11}. The 
availability of this last suite of quantum chemistry techniques however is rather limited at the moment,  
although it is likely to improve over the next few years~\cite{booth13}.

\subsection{Focus and organization of this review}
\label{subsec:focus}
The focus of the present critical review is on the assessment of the accuracy of DFT 
methods in the prediction and understanding of nanostructured GAM. Despite that the number 
of DFT studies published so far on GAM applications is enormous, there is not yet a 
well-established set of best practice principles for meaningful simulation of gas-adsorption 
phenomena and materials. The main objective of this review is to assist in defining this, by 
critically discussing key contributions and unsolved controversies in the field. Also, we 
identify here the subtle problems found in the interpretation of complex benchmark studies
and propose likely solutions for them.      

Our analysis is constrained to the design of carbon-based nanomaterials (CN) and MOF 
for applications in H$_{2}$ storage and CO$_{2}$ sequestration and capture. The main
reason justifying this choice is that these types of GAM and problems currently are attracting 
the highest attention within the community of computational and environmental materials 
scientists~\cite{mendoza12,fracaroli14,furukawa09,broom13,urbonaite08,tranchemontagne12,yurum09,dietzel09,assfour11,lan10,fang14}.
Moreover, by understanding the processes involved in the adsorption of H$_{2}$ and CO$_{2}$ 
molecules on CN and MOF we may infer also those affecting many other substances and materials.  
For instance, under normal conditions H$_{2}$ and CO$_{2}$ molecules are both linear  
and consequently cannot sustain any permanent dipole moment (e.g., like N$_{2})$; 
the predominant electrostatic interactions with the solid sorbent then are of hydrogen bond and 
quadrupolar type, respectively. However, when CO$_{2}$ molecules receive or donate electrons 
they normally get distorted and become polar. Thereby, from an electrostatic point of view 
CO$_{2}$ can turn out to be similar to other important molecules like, for instance, water and 
amonia. On the other hand, CN and MOF are structurally and electronically alike to boron-nitride 
nanostructures and covalent organic frameworks (see Fig.~\ref{figintro-1}), two families of GAM 
which at present are being investigated also very thoroughly~\cite{lan10b,feng12,reddy10,weng13,lin12}. 
For the sake of focus, other important mechanisms exploited in gas storage and sequestration processes 
(e.g., gas binding in bulk chemical form and thermodynamic destabilization of nanocrystals and 
clusters~\cite{seayad04,shevlin09,cazorla-amoros91,dean11,wang11}) 
and related families of materials (e.g., metal oxides and binary and ternary hydride 
compounds~\cite{ramachandran07,shevlin12,shevlin13,sun10b,gebhardt14,berube07}), will not be discussed 
exhaustively in the present work.

The organization of this review is as follows. In the next section we briefly summarize 
the binding energy targets pursued in the design of H$_{2}$ storage and CO$_{2}$ capture  
and sequestration GAM, together with a short description of CN and MOF.
Next, we explain the generalities of DFT methods and provide essential insight into its most popular
versions. In Sec.~\ref{sec:assessmentDFT}, we review the most recent and relevant works done in 
the assessment of the performance of DFT methods in GAM modeling. From them, we draw general 
conclusions on which DFT exchange-correlation functionals can be employed \emph{safely} for 
the simulation of gas-adsorption processes and materials. Also, we comment on the subtle 
problems found in the generalization of benchmark conclusions obtained in model cluster systems
to real GAM, and discuss possible solutions for them. In Sec.~\ref{sec:discussion}, we present a discussion 
on the current tendencies followed in first-principles design of GAM and propose a number of new 
directions to explore in prospective modeling and benchmark studies. Finally, our main conclusions 
are outlined in Sec.~\ref{sec:conclusions}.

\section{Nanostructured GAM}
\label{sec:matdescrip}
Nanostructured gas-adsorbent materials contain characteristic nanoscale motifs which 
are periodically repeated along one, two or three directions. Those motifs generally 
are composed of light atoms which congregate to form atomically sparse complexes. 
There are excellent review articles in the literature focusing on the physical properties 
and prospective implementations of GAM (see for instance Refs.~\cite{hu06,li11,orimo03,kuppler09,tasis06,pakdel11,nag10,park12}), 
hence we highlight here only the main traits of some representative species 
(i.e., CN and MOF). Also, we explain the basic requirements that potential GAM must accomplish 
for achieving effective storage and sequestration of H$_{2}$ and CO$_{2}$ gases. 

\subsection{GAM desiderata}
\label{subsec:desiderata}
The ``hydrogen storage'' and ``carbon dioxide capture and sequestration''  
problems have sparked very intense research within the communities of chemists, physicists
and engineers in the last past decade. The discovery of new materials with large
gas uptake capacities, robust thermodynamic stability, fast adsorption-desorption kinetics, 
and affordable production costs, is a key notion to succeed in the encountered gas storage 
and sequestration challenges~\cite{murray09,yang10,alessandro10,samanta12,choi09}.

The binding affinities of potential GAM are determined by their interactions with the       
gas molecules, which can be of electrostatic, dispersion, and/or orbital types~\cite{lochan06}.
In turn, the strength of the gas-GAM interactions depend on the characteristics of  
the materials and gas species, which in the latter case include electronic structure, 
atomic shape, polarizability and permanent dipole and/or quadrupole moments. 
Key quantities in the assessment of gas-adsorbent materials include
the alignment between frontier molecular orbitals (i.e., the energy difference between 
the highest occupied molecular orbital -HOMO- in the adsorbate and the lowest unoccupied
molecular orbital -LUMO- in the adsorbent), and the resulting charge transfers and 
binding energies. Let us have a closer look into the most relevant physical aspects of 
promising hydrogen and carbon capture GAM.

\subsubsection{H$_{2}$ storage}
\label{subsec:desiderata-h2}
For hydrogen to become the fuel of choice in future environmentally clean vehicles 
and electricity production plants, large amounts of H$_{2}$ need to be stored within 
relatively small volumes entailing moderate weights (for a detailed list of related 
targets see Refs.~\cite{doe1,doe2}). In this context, H$_{2}$ adsorption on light solid 
sorbents through weak and mild molecule-GAM interactions emerges as one of the most 
promising routes. (Dissociative adsorption in solid metal hydrides where molecular 
hydrogen is dissociated to form a solid solution with the hydride, is also considered as 
propitious~\cite{seayad04,shevlin09}; however, we will not discuss such an interesting 
gas-storage mechanism or related family of materials in the present review for the 
sake of focus.) 
Current estimations of the optimal binding energy for adsorption of hydrogen   
at ambient temperature and considering safe delivery pressure conditions of $1.5-30$~bar,
amount to $\sim 0.2$~eV/molecule~\cite{garberoglio05,bathia06,basile14}. 
Meanwhile, kinetic effects may be critical for practical applications (e.g., complete 
H$_{2}$ storage-release cycles must be realized within seconds) and when these are taken into 
account favourable binding energies turn out to be $\sim 0.7$~eV/molecule~\cite{li03}. 
Based on these assessments, the binding energy range that typically is targeted in 
most \emph{ab initio} H$_{2}$-storage works is $0.1-1.0$~eV/H$_{2}$, which spans from  
dispersion to moderately covalent molecule-GAM interactions (see Table~I). 

The strength of the H$_{2}$-solid sorbent interactions can be tuned by decorating the  
GAM surfaces with transition or alkaline-earth metal atoms, and this normally tends to  
increase the affinity towards gas binding~\cite{bhattacharya10,hamaed10,phillips08}. 
The orbital mechanism acting behind this effect is the Kubas interaction, which 
consists in electron donation from the H$_{2}$ $\sigma$-bonding orbital to the 
empty metal $d$ orbitals and simultaneous electron back-donation from the filled metal
$d$ orbitals to the H$_{2}$ $\sigma$-antibonding orbital~\cite{kubas01,kubas07}.
A different approach that is promising for gas storage at elevated temperatures,
is the spillover of hydrogen onto inert surfaces. This mechanism implies the chemical 
activation of H$_{2}$ molecules placed on top of transition metal sites through 
catalysis, followed by the migration of atomic hydrogen onto the surface of the receptor 
material (e.g., activated carbons, COF, and MOF). Record hydrogen storage capacities 
of $4-6$~wt.~\% have been recently accomplished with this strategy in metal-carbon complexes 
and MOF~\cite{tsao10,psofogiannakis09,psofogiannakis11,li07,wang12b}.  

\begin{figure*}
\centerline{
\includegraphics[width=0.90\linewidth]{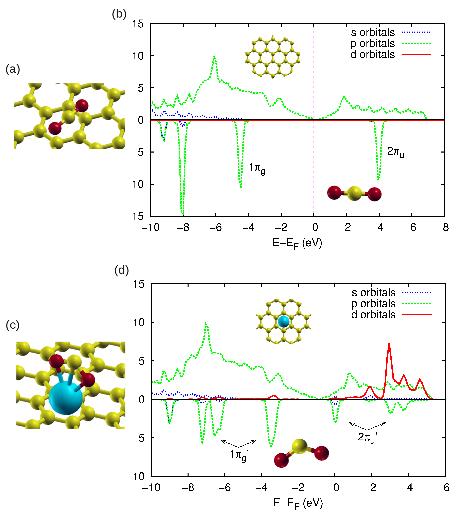}}
\vspace{-0.50cm}
\caption{(a)~Sketch of a carbon dioxide molecule interacting with a carbon-based
        surface. (b)~Partial density of electronic states calculated in the system
        sketched in (a), expressed in arbitrary units. The electronic states 
        localized in the carbon surface are represented in the upper part of the
        figure, and those corresponding to the gas molecule in the lower.
        (c)~Sketch of a carbon dioxide molecule interacting with a Ca-decorated
        carbon surface. (d)~Partial density of electronic states calculated in the 
        system sketched in (c), expressed in arbitrary units. The electronic states    
        localized in the Ca-decorated carbon surface are represented in the upper 
        part of the figure and those corresponding to the gas molecule in the lower. 
        HOMO bonding and LUMO anti-bonding molecular orbitals are indicated in 
        panels (b) and (d) [see text].}
\label{fig:kubas-co2}
\end{figure*}

\subsubsection{Carbon capture and sequestration}
\label{subsec:desiderata-ccs}
Carbon capture and sequestration (CCS) actions are being implemented in fossil-fuel 
burning power plants for cutting the amounts of green-house gases which are expelled 
to the atmosphere. Current CCS means mostly rely on the scrubbing of synthesis and flue 
gases with amine solvents. However, solvent-based CCS technologies generally are not 
cost-effective due to the high energy penalty involved in solvent regeneration and 
equipment corrosion~\cite{aaron05}. Fortunately, membranes and solid sorbents do not 
suffer from these important drawbacks and thus they constitute the core of next-generation 
CCS technologies. 

There are two main types of CCS processes in which GAM are highly promising: pre-combustion and 
post-combustion. In pre-combustion CCS, the fuel is converted into a mixture of hydrogen and carbon 
dioxide gases using processes such as ``gasification'' or ``reforming''. The usual thermodynamic 
conditions in pre-combustion CCS are high temperatures and high pressures (i.e., above 100$^{\circ}$ 
and $20-30$~bar, respectively), and the concentration of CO$_{2}$ gas is high. At those conditions, 
it can be shown that CO$_{2}$ sequestration turns out to be effective when the strength of the gas-GAM 
interactions roughly amounts to $2.0$~eV/molecule. For this, consider the inverse calcination 
reaction, ${\rm CaO} + {\rm CO_{2}} \to {\rm CaCO_{3}}$, to be a model of a typical pre-combustion 
CCS process~\cite{cazorla-amoros91,dean11,wang11}. The Gibbs free-energy balance of this reaction has been 
approximated by the function $\Delta G = -1.83 + 0.0016 \cdot T$~eV/molecule, where $T$ represents 
the temperature (expressed in Kelvin)~\cite{gilchrist89}. Now, by assuming that the thermal contributions 
to $\Delta G$ mostly arise from the entropy of the gas, $E_{\rm bind} \sim 2.0$~eV/molecule follows. 

In post-combustion CCS, the CO$_{2}$ gas is separated from the exhaust of a combustion process. 
The usual thermodynamic conditions in post-combustion CCS are low temperatures and low partial 
CO$_{2}$ pressures. In that case, it can be shown that the ideal gas-binding energy scale for  
solid sorbents is $\sim 0.2$~eV/molecule. For this, consider that a typical CCS post-combustion 
process can be described by the generic capture reaction ${\rm A} + n{\rm CO_{2}} \to {\rm A (CO_{2})}_{n}$, 
and that the corresponding Gibbs free-energy balance vanishes at temperatures close to ambient (e.g., 
$50^{\circ}$ degrees Celsius). Now, by assuming that the entropy change in the solid sorbent ${\rm A}$ 
upon adsorption of molecules is practically null and using tabulated thermodynamic data of the 
CO$_{2}$ gas (i.e., the corresponding entropy is $S_{\rm gas} = 5.3 \cdot 10^{-4}$~eV/molecule~K), 
$E_{\rm bind} \sim 0.2$~eV/molecule follows. 

Based on these assessments, typical gas-adsorption energies pursued in most \emph{ab initio} CCS 
studies span from $0.2$ to $2.0$~eV/CO$_{2}$, and depending on the specific aims one end or the 
other of this interval is targeted. As it is shown in Table~I, that energy range extends from 
weak dispersion to strong covalent molecule-GAM interactions. In addition to these binding energy 
requirements, prospective CCS GAM need to display also favorable selectivity features with respect 
to the adsorption of N$_{2}$, oxygen and water, since those species are also abundant in the generated 
flue and synthesis gases.

It is worth mentioning that similar electronic orbital processes to the Kubas interaction~\cite{kubas01,kubas07} 
explained in the previous section, can be exploited also for enhancing the affinities of GAM towards CO$_{2}$ 
binding. In Fig.~\ref{fig:kubas-co2}, we represent the partial density of electronic states 
calculated for a carbon dioxide molecule interacting with a pristine carbon surface, and with 
the same surface decorated with Ca atoms. The presence of calcium dopants induces overlappings 
between electronic $p$ CO$_{2}$ orbitals and $s,d$ metallic states in the region surrounding 
the Fermi level, resulting in a transfer of charge from the molecule to the metal centers. In 
this process the degeneracy of the HOMO $1\pi_{g}$ bonding and LUMO $2\pi_{u}$ anti-bonding 
molecular orbitals is lifted, and electronic charge is back-donated from the filled $s,d$ Ca 
orbitals to the anti-bonding CO$_{2}$ electronic states (compare bottom panels in both Figs.~\ref{figintro-3} 
and~\ref{fig:kubas-co2}). This orbital mechanism provokes the bending of the gas molecule 
and intensifies the gas-GAM interactions~\cite{cazorla11}, thus providing a route for the rational 
design of sorbent materials for pre-combustion CCS applications.

\subsection{Carbon nanomaterials}
\label{subsec:c-b-n}
Carbon nanostructures (CN) include an ample range of carbon allotropes such as 
nanotubes, graphene, met-cars and graphite intercalation compounds (see Fig.~\ref{figintro-1}). 
CN exhibit a large variety of electronic and transport 
properties resulting from the specific way in which atoms are arranged.
Carbon nanotubes, for instance, may be metallic or semiconducting depending
on their diameter and the degree of helicity present in their hexagonal-ring walls~\cite{wilder98,castro09}. 
Met-cars (i.e., metallocarbohedrynes) are extremely stable symmetric clusters 
formed by metal and carbon atoms with stoichiometric formula $M_{n}$C$_{m}$, 
where $M$ typically stands for a transition metal atom ($M =$ Ti, V, Zr, Hf, 
Fe, Cr, and Mo) and $n, m = 8, 12$ (see Fig.~\ref{figintro-1})~\cite{rohmer93,rohmer95}. 
Graphite intercalation compounds (not shown in Fig.~\ref{figintro-1}) are composed of 
alternating planes of transition, alkali or alkaline-earth metal and C atoms disposed in 
triangular and hexagonal lattices, respectively. In these materials relative shifts
between successive atomic planes along the out-of-plane direction may occur leading 
to the formation of stacking patterns~\cite{dresselhaus02,emery05,cobian08}. 

Interestingly, the electronic and gas-adsorption properties of CN can be finely tuned by 
coating their surfaces with dopant species, creating defects, or imposing mechanical 
strains~\cite{lee10,cazorla10,krasheninnikov07,dresselhaus10,zhou10,dutta14}. 
Among these, the doping strategy with metal adatoms has attracted a lot of attention 
because of its technical simplicity and expected enhancement of the binding affinity 
towards gas molecules through the Kubas interaction~\cite{kubas01,kubas07} 
(see Sec.~\ref{subsec:desiderata-h2})~\cite{shevlin08,yildirim05,ataca09,cazorla12}.
However, transition metal atoms in carbon surfaces exhibit a strong tendency towards 
clustering~\cite{sun05,li06,cazorla09} and thereby further developments in the field of 
synthetic chemistry addressing this problem are awaited. Alternatively, the creation of 
nanostructured networks of defect sites such as vacancies, nanoribbons and islands, may 
increase considerably the number of unsaturated carbon sites. This kind of structural 
modifications generally enhance the interactions with the gas molecules and therefore 
lead to improved gas storage capacities~\cite{wang13,banhart10}.

\subsection{Metal-organic frameworks}
\label{subsec:mofs}
Metal-organic frameworks (MOF) are multi-dimensional nanoporous structures 
composed of metal ions (or clusters) coordinated to rigid organic molecules, which are called 
linkers. The choice of the metal ion and linker species completely determines the structure and 
functionality of the resulting MOF. Common organic linkers include bidentate carboxylics 
(e.g., HOOC-COOH), tridendate carboxylates (e.g., C$_{9}$H$_{6}$O$_{6}$), 1,4-benzenedicarboxyalte
(BDC), and azoles (e.g., C$_{2}$H$_{3}$N$_{3}$) molecules. The great freedom with which the linkers 
and metal ions can be chosen and combined is reflected in the more than $20,000$ MOF species that have 
been reported in the last two decades (for an extensive review on this topic, see Ref.~\cite{furukawa13}). 
Among these, the $A$-MOF74 ($A =$ Mg and Ni), MOF-5 (e.g., Zn$_{4}$O tetrahedra linked by BDC 
organic molecules) and $X$-BTT ($X =$ Ca, Fe, Mn,Cu, Co, Ni, Cr, and Cd, and BTT = 1, 2, 
5-benzenetristetrazolate) families emerge as three of the most investigated compounds. 
The $A$-MOF74 structure is based on coordinated carboxyl and hydroxy groups (i.e., helical $A$-O-C 
rods that emanate from 6-coordinated $A$ centers) and its primitive unit cell contains $54$ atoms. 
The MOF-5 crystal has a cubic symmetry and its conventional and primitive cell contain eight and 
four OZn$_{4}$O(BDC)$_{3}$ formula units, respectively (i.e., $424$ and $106$ atoms).  
The primary building block of X-BTT is a truncated octahedron (i.e., a six [X$_{4}$Cl]$^{+7}$ squares 
and eight BTT ligands structure) which shares its square faces to form a general cubic framework, 
comprising a total of $210$ atoms. 

The polarity of the building units and spatial separation between the organic linkers, affect 
profoundly the binding strength of H$_{2}$ and CO$_{2}$ molecules and thereby constitutes a rationale 
for the design of MOF-based GAM. Also, decorating the MOF surfaces with alkali, alkaline-earth 
and transition metal atoms can enhance significantly the adsorption of gas species. Practical 
realizations of this last type of functionalization, however, remain yet technically 
difficult due to potential atomic coalescence~\cite{kaye08,schroder08,dixit12}. 
Alternatively, modifications of the porous frameworks based on the embedment of metal nanoparticles 
have been recently investigated, producing very impressive gas-capacity records~\cite{cheon09,lim12,jeon11}.

\section{Density Functional Theory Methods}
\label{sec:dft}

\begin{table*}
\begin{center}
\label{tab:physvschem}
\begin{tabular}{c c c c c c}
\hline
\hline
$ $ & $ $ & $ $ & $ $ & $ $ & $ $ \\
$\quad {\rm Interaction~type}~\to \quad $ & $\quad {\rm Covalent} \quad$ & $\quad {\rm Ionic} \quad$ & $\quad {\rm Hydrogen~bond} \quad$ & $\quad {\rm Dispersion} \quad$ & $\quad {\rm Many-body}\quad$ \\
$\quad {\rm Aspect} \quad $ & $ $ & $ $ & $ $ & $ $ & $ $ \\
$ \downarrow $ & $ $ & $ $ & $ $ & $ $ & $ $ \\
\hline
$ $ & $ $ & $ $ & $ $ & $ $ & $ $ \\
${\rm Spatial~decay}$ & $\exp{[-r^{2}]}$ & $\frac{1}{r}$ & $\frac{1}{r^{3}}$ & $\frac{1}{r^{N}}~(N \ge 6) $ & ${\rm System~dependent}$ \\
$ $ & $ $ & $ $ & $ $ & $ $ & $ $ \\
${\rm Energy~scale~(eV/molecule)}$ & $1.0-0.1$ & $1.0-0.1$ & $0.1$ & $0.1-0.01$ & $0.1-0.001$ \\
$ $ & $ $ & $ $ & $ $ & $ $ & $ $ \\
\hline
\hline
\end{tabular}
\end{center}
\caption{Characteristic traits of usual bonded and nonbonded atomic interactions taking place in 
         condensed matter systems and surfaces.}
\end{table*}

\subsection{General considerations}
\label{subsec:general}
The wave function of a $N$-electron system, $\Psi ({\bf r}_{1}, {\bf r}_{2},...,{\bf r}_{N})$, 
contains all its physical information and it is determined by solving the corresponding 
Schr\"{o}dinger equation. 
In real materials electrons interact through the Couloumb repulsion and are abundant,
thus $\Psi$ is a complex mathematical function that in most of the cases is unknown.
In 1965, Kohn-Sham developed an ingenious theory to effectively calculate the energy and
related properties of many-electron systems without the need of explicitly knowing
$\Psi$~\cite{kohn65,sham66}. The main idea underlying this theory, called density functional 
theory (DFT), is that the exact ground-state energy, $E$, and electron density,
$n({\bf r})$, of a many-electron system can be determined by solving an effective one-electron 
Schr\"odinger equation of the form:
\begin{equation}
{\rm H}_{eff} \psi_{i \sigma} = \epsilon_{i \sigma} \psi_{i \sigma}~, 
\label{eq:onelectron}
\end{equation}
where index $i$ labels different one-electron orbitals and $\sigma$ the corresponding spin state.
In particular,  
\begin{equation}
{\rm H}_{eff} = -\frac{1}{2}\nabla^{2} + V_{ext}({\bf r}) + \int \frac{n({\bf r'})}{|{\bf r} - {\bf r'}|} d{\bf r'} + V_{xc}({\bf r})~,
\label{eq:heff}
\end{equation}
and
\begin{equation}
n({\bf r}) = \sum_{i \sigma} |\psi_{i \sigma} ({\bf r})|^{2}~,
\label{eq:density}
\end{equation}
where $V_{ext}$ represents an external field and $V_{xc} ({\bf r}) = \delta E_{xc} / \delta n ({\bf r})$
is the potential function that results from deriving the exchange-correlation energy, $E_{xc}$, with respect 
to the electron density. 

The exchange-correlation energy has a purely quantum mechanical origin and can be defined as 
the interaction energy difference between a many-electron quantum system and its classical counterpart. 
Electrons are indistinguishable particles called fermions and the wave function describing an ensemble 
of electrons must change its sign when two particles exchange orbital states. This quantum antisymmetry 
leads to an effective repulsion between electrons, called the Pauli repulsion, which helps in lowering 
their total Coulomb energy. Despite $E_{xc}$ represents a relatively small fraction of the total energy, 
this is an extremely crucial quantity for all the physical aspects of materials and molecules because 
it acts directly on the bonding of atoms. The $E_{xc} [n]$ functional generally is unknown and in 
practice needs to be approximated. Actually, this is the only source of fundamental error in DFT methods 
and depending on the approximation that is used the resulting approach may turn out to be valid or not 
for describing the systems and phenomena of interest.

In standard cases $E_{xc} [n]$ is approximated with the expression
\begin{equation}
E_{xc}^{approx}[n] = \int \epsilon_{xc}^{approx}({\bf r}) n({\bf r}) d{\bf r}~,
\label{eq:excapprox}
\end{equation}
where $\epsilon_{xc}^{approx}$ is made to depend on $n({\bf r})$, $\nabla n({\bf r})$, 
and/or the electronic kinetic energy 
$\tau ({\bf r}) = \frac{1}{2} \sum_{i \sigma} |\nabla \psi_{i \sigma} ({\bf r})|^{2}$
(see next sections). 
Actually, the exact form of the exchange-correlation energy is readily known and reads 
\begin{equation}
E_{xc} [n] = \int  n({\bf r}) d{\bf r}  \int \frac{n_{xc}({\bf r} , {\bf r'})}{|{\bf r} - {\bf r'}|} d{\bf r'}~,
\label{eq:exc}
\end{equation}
where $n_{xc}({\bf r} , {\bf r'}) = n_{x} ({\bf r} , {\bf r'}) + n_{c} ({\bf r} , {\bf r'})$ 
is the exchange-correlation hole density at position ${\bf r'}$ that surrounds an electron at 
position {\bf r}. Some important constraints on $n_{xc}({\bf r} , {\bf r'})$ have already
been established. For instance, $n_{x} ({\bf r} , {\bf r'})$ must be nonpositive everywhere and 
its space integral is equal to $-1$. Also, the space integral of the correlation hole density is 
zero. These constraints can be employed in the construction of approximate, but physically 
correct, $E_{xc} [n]$ functionals.

\begin{figure}
\centerline{
\includegraphics[width=1.00\linewidth]{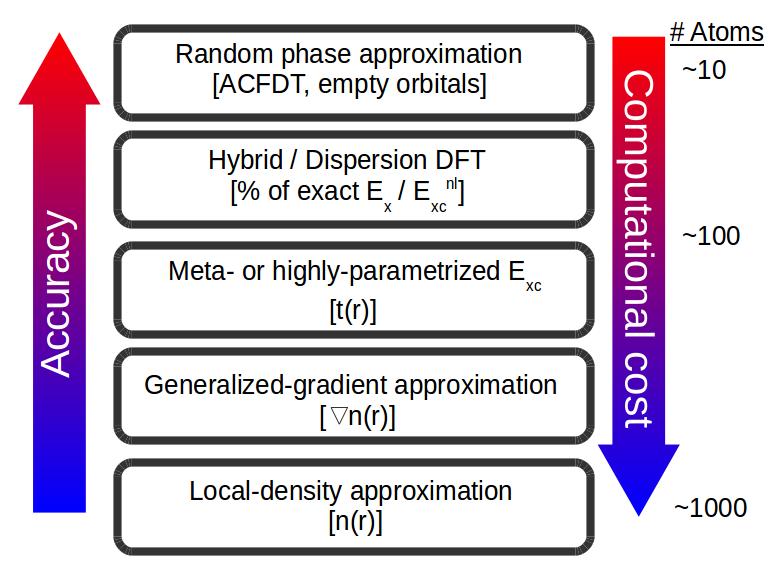}}
\caption{Representation of common levels of $E_{xc}$ approximation within density functional 
         theory together with some general features. Regions coloured in red indicate ``High'' and
         in blue ``Low''. The typical size of the systems that can be handled with those
         approaches are indicated on the right margin of the figure.}
\label{fig4}
\end{figure}

In the next subsections, we review the most popular $E_{xc} [n]$ approximations which 
are currently employed in computational studies of GAM. In Table~I, we summarize the 
main types of bonded and nonbonded atomic interactions taking place in condensed matter 
systems and surfaces. In Table~II, we outline the adequacy of the considered DFT methods 
in describing those interactions. In Table~III, we list some popular computer simulation 
packages that can be used to perform standard and more advanced DFT calculations. Finally, 
the relative degree of accuracy and computational expense associated to those DFT 
approaches are sketched in Fig.~\ref{fig4}. Additional details on these topics can be 
found in some recent specialized reviews~\cite{perdew13,klimes12,scheffer12,dobson12,steinmann12}.

\subsection{Local and semi-Local $E_{xc}$ energy functionals}
\label{subsec:local}
In local approaches (e.g., local density approximation -LDA-), $E_{xc}$ is approximated 
with Eq.~(\ref{eq:excapprox}) and the exchange-correlation energy is taken to the be  
that of an uniform electron gas with density $n ({\bf r})$, namely $\epsilon_{xc}^{unif}$.  
The exact expression of the $\epsilon_{xc}^{unif} [n]$ functional is known 
numerically from accurate quantum Monte Carlo calculations~\cite{perdew81,ceperley80}. 
In order to deal with the nonuniformity of realistic many-electron systems, these are normally 
partitioned into infinitesimal volume elements which are considered to be locally uniform.  
In semilocal approaches (e.g., generalized gradient approximation -GGA-), $E_{xc}$ is 
approximated also with Eq.~(\ref{eq:excapprox}) and $\epsilon_{xc}^{approx}$ is made to 
depend on $n({\bf r})$ and its gradient $\nabla n({\bf r})$~\cite{pw91,pbe}.

Both local and semilocal approximations satisfy certain exact $E_{xc}$ constraints 
(e.g., some exact scaling relations and the exchange-correlation hole sum rules) 
and can work notably well for systems in which the electronic density varies slowly 
over space (e.g., bulk crystals at equilibrium conditions, see Table~II). However, 
this is manifestly not the case of GAM which normally contain surfaces and pores 
where $n({\bf r})$ can change very abruptly.  
Moreover, by construction local and semilocal functionals totally neglect   
long-range electron correlations, otherwise known as dispersion interactions, 
which certainly are ubiquitous in gas-adsorption phenomena. 

\subsection{Meta-GGA or highly parametrized $E_{xc}$ energy functionals}
\label{subsec:meta}
Meta-GGA functionals are semilocal in nature (i.e., assume the approximate $E_{xc} [n]$ 
expression in Eq.~(\ref{eq:excapprox}) but contain an additional degree of elaboration 
with respect to LDA and GGA: the orbital kinetic energy density, 
$\tau ({\bf r}) = \frac{1}{2} \sum_{i \sigma} |\nabla \psi_{i \sigma} ({\bf r})|^{2}$. 
This new ingredient allows one to construct functionals which satisfy some additional 
exact constraints, like for instance the correct gradient expansion of the exchange energy 
up to fourth order. Another interesting feature of this family of functionals is that 
they can be trained to capture the short- and middle-range parts of electronic correlation, 
which in some special cases may be enough to describe correctly the binding of atoms 
(see Table~II).

Meta-GGA functionals are versatile and in general do not entail significantly larger 
computational expense than standard LDA and GGA. Examples of this family include the suite 
of TPSS functionals due to Perdew, Scuseria, and others~\cite{revtpss,vdwrevtpss}, and 
the ``Minnesota functionals'' (M$X$, with $X =$ 05, 06, 08, 11, and 12) developed by 
Truhlar and his group in the University of Minnesota~\cite{truhlar}. 
Meta-GGA functionals in general can provide accurate lattice constants in solids, surface 
energies, and molecular binding energies, and are already implemented in popular quantum 
chemistry and DFT packages like, for instance, GAUSSIAN~\cite{gaussian}, NWCHEM~\cite{nwchem}, 
SIESTA~\cite{siesta}, and VASP~\cite{vasp} (see Table~III). Nevertheless, dispersion interactions 
cannot be reproduced systematically with these functionals since they lack of explicit nonlocal 
correlation kernels (see Sec.~\ref{subsec:vdw} below).

\subsection{Hybrid exchange energy functionals}
\label{subsec:hybrid}
Hybrid functionals comprise a combination of nonlocal exact Hartree-Fock (HF) and local 
exchange energies, together with semilocal correlation energies. The proportion in which both 
nonlocal and local exchange densities are mixed generally relies on empirical rules. 
The popular B3LYP~\cite{b3lyp}, for instance, takes a $20$~\% of the exact HF exchange energy 
and the rest from the approximate GGA and LDA functionals. Other well-known hybrid functionals 
are the HSE proposed by Heyd-Scuseria-Ernzerhof~\cite{hse}, PBE0~\cite{pbe0}, and the family 
of Minnesota meta hybrid GGA~\cite{truhlar}.

In contrast to local and semilocal functionals, hybrids describe to some extent the 
delocalization of the exchange-correlation hole around an electron. This situation becomes 
of particular relevance in chemisorption and charge transfer processes, where the atomic 
bonds turn out to be elongated or shortened. Also, in strongly correlated systems containing $d$ 
and $f$ electronic orbitals life, for instance, transition metal oxides. In other words, hybrid 
functionals are useful in situations where the electron self-interaction errors, stemming 
from an imperfect cancellation between the artificial interaction of an electron with itself 
and the exchange energy, are potentially large. Hybrid functionals, however, do not account for 
the long range part of the correlation hole energy and thus cannot reproduce dispersion forces  
(see Table~II). Efforts to effectively correct for such drawbacks have been made recently 
by Head-Gordon and other authors (e.g., the range separated and dispersion corrected 
$\omega$B97X-D and $\omega$M06-D3 functionals)~\cite{gordon08,lin13,schwabe07,mardirossian14}. 

\subsection{Dispersion-corrected $E_{xc}$ energy functionals}
\label{subsec:vdw}
The condition that any DFT-based dispersion scheme must accomplish is to reasonably reproduce
the asymptotic $1/r^{6}$ behaviour of the interaction between two particles separated by a 
distance $r$ in a gas. Local, semilocal, and hybrid energy functionals totally miss 
this requirement. The most straightforward way to correct for such a fault consists in 
adding an energy term to the exchange-correlation energy that is attractive and has the form 
$E_{\rm disp} = -\sum_{i,j} C_{ij} / r^{6}_{ij}$ (where indexes $i$ and $j$ label different 
particles, and a damping factor is introduced at short distances in order to avoid divergences). 
This approximation represents the core of a suite of methods termed DFT-D which, due to their 
simplicity and low computational cost, are being widely used at the moment. Probably, the most 
popular family of DFT-D methods are the dispersion corrected GGA functionals proposed by Grimme~\cite{grimme}. 
Despite their appealing features, DFT-D methods present several shortcomings. First, many-body 
dispersion effects and faster decaying terms such as the $B_{ij} / r^{8}_{ij}$ and $C_{ij} / r^{10}_{ij}$ 
interactions are completely disregarded. Second, it is not totally clear from where one should 
obtain the optimal $C_{ij}$ coefficients. And finally, once these parameters are determined 
they remain fixed during the simulations, and this can be problematic in situations where the 
orbital hybridization and oxidation states change as compared to the free atoms case.  

Several improvements on DFT-D methods have been proposed, in which the value of the
dispersion coefficients are made to depend on the specific atomic environment. 
Those correspond to the DFT-D3 method by Grimme~\cite{grimme-d3}, the vdW(TS) approach by 
Tkatchenko and Scheffler~\cite{ts}, and the BJ model by Becke and Johnson~\cite{bj}. 
In those approaches the specific variation of the $C_{ij}$ parameters are taken on the atomic 
coordination or effective volume, and the calculation of reference dispersion coefficients and 
atomic polarizabilities are required. 

A third degree of elaboration exists in which no external input parameters 
are needed and the dispersion interaction is directly computed from the electron density. 
In this context, the exchange-correlation energy is expressed as 
$E_{xc} = E_{x}^{\rm GGA} +  E_{c}^{\rm LDA} + E_{c}^{\rm nl}$ where 
$E_{c}^{\rm nl}$ is the nonlocal correlation energy. Particularly, $E_{c}^{\rm nl}$ is 
calculated with a double space integral involving the electron density and a two-position 
integration kernel, $\phi$, of the form 
\begin{equation}
E_{c}^{\rm nl} [n] = \frac{1}{2} \iint n({\bf r'}) \phi(q',q, |{\bf r'} -{\bf r}|) n({\bf r}) d{\bf r'} d{\bf r}~.
\label{eq:nlc}
\end{equation}
In Eq.~(\ref{eq:nlc}), $q'$ and $q$ are the values of an universal function evaluated in positions 
${\bf r'}$ and ${\bf r}$, and $\phi$ is a complex function that obeys two main constraints: 
$E_{c}^{\rm nl}$ is strictly zero for any system with constant $n$, and the interaction between
two molecules has the correct $|{\bf r'} -{\bf r}|^{-6}$ dependence for long distances.
The described approach was introduced by Dion \emph{et al.} in 2004~\cite{dion04} 
and it represents a key DFT development since it combines all types of interaction 
ranges within a same formula. Refinements on Dion's approach have subsequently appeared where 
(i)~the original two-position integration kernel $\phi$ is modified (e.g., the so-called nonlocal VV10 
functional due to Vydrov and Voorhis~\cite{vydrov10,vydrov12,sabatini13}), and (ii)~the 
exchange term in $E_{xc}$ is substituted with other more accurate functionals 
(e.g., the so-called vdW-DF2~\cite{lee10b}, vdW-optB88 and vdW-optPBE~\cite{optex}, and 
vdW-C09$_{x}$~\cite{cooper10} schemes). All these approaches are termed vdW-DF and, thanks 
to the seminal work of Rom\'{a}n-P\'{e}rez and Soler~\cite{roman09}, their accompanying 
computational expense in a planewave framework is moderate in practice. Nevertheless, the way 
in which nonlocal correlations are calculated inherently assumes pairwise additivity and thus 
many-body effects are completely disregarded in vdW-DF methods (see Table~II). 

\begin{table*}
\begin{center}
\label{tab:dftfunct}
\begin{tabular}{c c c c c c}
\hline
\hline
$ $ & $ $ & $ $ & $ $ & $ $ & $ $ \\
$\quad {\rm Interaction~type}~\to \quad $ & $\quad {\rm Covalent} \quad$ & $\quad {\rm Ionic} \quad$ & $\quad {\rm Hydrogen~bond} \quad$ & $\quad {\rm Dispersion} \quad$ & $\quad {\rm Many-body}\quad$ \\
$\quad {\rm DFT~flavour} \quad $ & $ $ & $ $ & $ $ & $ $ & $ $ \\
$ \downarrow $ & $ $ & $ $ & $ $ & $ $ & $ $ \\
\hline
$ $ & $ $ & $ $ & $ $ & $ $ & $ $  \\
${\rm Local~and~semi-local}$ & $\surd$ & $\surd$ & $\surd/\times$ & $\times$ & $\times$ \\
${\rm (LDA,GGA)} $ & $ $ & $ $ & $ $ & $ $ & $ $ \\
\cite{perdew81,ceperley80,pw91,pbe} & $ $ & $ $ & $ $ & $ $ & $ $ \\
$ $ & $ $ & $ $ & $ $ & $ $ & $ $ \\
${\rm Highly~parametrized}$ & $\surd$ & $\surd$ & $\surd/\times$ & $\surd/\times$ & $\times$ \\
${\rm (Meta-GGA,Minnesota)} $ & $ $ & $ $ & $ $ & $ $ & $ $ \\
\cite{revtpss,vdwrevtpss,truhlar} & $ $ & $ $ & $ $ & $ $ & $ $ \\
$ $ & $ $ & $ $ & $ $ & $ $ & $ $ \\
${\rm Hybrid}$ & $\surd$ & $\surd$ & $\surd$ & $\times$ & $\times$ \\
${\rm (B3LYP,HSE)} $ & $ $ & $ $ & $ $ & $ $ & $ $ \\
\cite{b3lyp,hse,pbe0} & $ $ & $ $ & $ $ & $ $ & $ $ \\
$ $ & $ $ & $ $ & $ $ & $ $ & $ $ \\
${\rm Dispersion-corrected}$ & $\surd$ & $\surd$ & $\surd$ & $\surd$ & $\times$ \\
${\rm (DFT-D2,vdW-DF,VV10)} $ & $ $ & $ $ & $ $ & $ $ & $ $ \\
\cite{grimme,dion04,optex,lee10b,vydrov10} & $ $ & $ $ & $ $ & $ $ & $ $ \\
$ $ & $ $ & $ $ & $ $ & $ $ & $ $ \\
${\rm Many-body}$ & $\surd$ & $\surd$ & $\surd$ & $\surd$ & $\surd$ \\
${\rm (DFT+MBD, RPA-DFT)} $ & $ $ & $ $ & $ $ & $ $ & $ $ \\
\cite{tkatchenko12b,ambrosetti14,scheffer12,dobson12} & $ $ & $ $ & $ $ & $ $ & $ $ \\
$ $ & $ $ & $ $ & $ $ & $ $ & $ $ \\
\hline
\hline
\end{tabular}
\end{center}
\caption{Description of the performance of some DFT variants in describing usual types 
         of bonded and nonbonded interactions found in condensed matter systems and surfaces.
         Symbol $\surd$ ($\times$) indicates correct (incomplete) description of the 
         considered type of interaction by the corresponding DFT method.}
\end{table*}

\subsection{The random phase approximation and DFT+MBD}
\label{subsec:rpa}
The adiabatic connection fluctuation-dissipation (ACFD) theorem provides a general and exact 
expression for the exchange-correlation energy of a many-electron system, thereby $E_{xc}$ 
in principle can be calculated in a very accurate way incorporating higher-order many-body effects. 
In particular, the correlation energy of a system adopts the form 
\begin{equation}
\begin{split}
E_{c} = & -\frac{1}{2 \pi} \int_{0}^{\infty} d\omega \quad \times \\
    &  \int_{0}^{1} {\bf Tr}\left[ \frac{\chi_{\lambda} \left({\bf r'}, {\bf r}, i\omega \right) - \chi_{0} \left({\bf r'}, {\bf r}, i\omega \right)}{|{\bf r'} -{\bf r}|}\right] d\lambda~,
\end{split} 
\label{eq:acfd}
\end{equation}
where $\chi_{0} \left({\bf r'}, {\bf r}, i\omega \right)$ is the bare density-density response function, 
$\chi_{\lambda} \left({\bf r'}, {\bf r}, i\omega \right)$ the interacting density-density response function 
at Coulomb coupling strength $\lambda$, and ${\bf Tr}$ denotes the six-dimensional integration over the 
variables ${\bf r'}$ and ${\bf r}$. The response function $\chi$ measures the electronic response of the 
system at a point ${\bf r'}$ due to a frequency-dependent electric field perturbation at a point
${\bf r}$. In the ACFD approach, the adiabatic connection between a reference non-interacting system
(defined at $\lambda=0$) and the fully interacting system (defined at $\lambda=1$) provides the correlation 
energy of the latter, including many-body dispersion as well as other types of electron correlation effects.
This theoretical framework has been exploited by several authors to develop novel many-body DFT-based 
approaches, among which we highlight the random phase approximation to DFT and the DFT+MBD method. 
 
In the random phase approximation (RPA-DFT) scheme, the interacting response function $\chi_{\lambda}$
is defined self-consistently via the equation $\chi_{\lambda} = \chi_{0} + \lambda \frac{\chi_{0} \chi_{\lambda}}{|{\bf r'} -{\bf r}|}$.
This approximation has been shown to work reasonably well for a number of cluster and extended systems~\cite{ren11}. 
Following the Adler-Wiser formalism~\cite{adler62,wiser63}, $\chi_{0}$ can be computed by using the occupied 
and virtual orbitals, and the corresponding energies and occupancies obtained in DFT calculations.
In analogy to post-HF approaches, the computational expense associated to RPA-DFT is very large (i.e., 
typically scales with the fourth power of the number of particles) and the convergence with respect 
to the basis set generally is too slow~\cite{lu09,eshuis10,eshuis11,eshuis12,umari10}. 
Also, it must be noted that the short-range part of the electron correlation energy is not precisely 
reproduced by RPA-DFT and that this shortcoming may be problematic in studying molecular systems~\cite{furche01}.

Meanwhile, recent efforts done in the groups of Tkatckenko and Scheffler have given birth to
the so-called DFT+MBD method~\cite{ruiz12,tkatchenko12b,ambrosetti14,distasio14}, which accounts 
also correctly for the Coulomb screening and many-body effects in electronic systems. In the DFT+MBD 
approach, the Schr\"{o}dinger equation of a set of fluctuating and interacting quantum harmonic 
oscillators is solved directly within the dipole approximation, and the resulting many-body energy is 
coupled to an approximate semilocal DFT functional. These approximations result in a significant reduction 
of computational load as compared to the RPA-DFT method, allowing one to describe larger systems 
containing up to few hundreds of atoms.  

The most appealing features of the RPA-DFT and DFT+MBD methods is that they are very accurate and 
suitable for studying small-gap and metallic periodic systems. In fact, these techniques 
have already been applied successfully to the study of strongly correlated crystals and surfaces~\cite{harl10,xiao12,schimka10}, 
organic molecules adsorbed on metallic surfaces~\cite{ruiz12}, and ionic and semiconductor 
solids~\cite{zhang11}. Effective schemes of the RPA-DFT and DFT+MBD methods have already been implemented 
in commercial and open-source first-principles packages like, for instance, ABINIT~\cite{abinit}, 
GPAW~\cite{gpaw}, and VASP~\cite{vasp} (see Table~III).

\begin{table*}
\begin{center}
\label{tab:codes}
\begin{tabular}{c c c c c c c}
\hline
\hline
$ $ & $ $ & $ $ & $ $ & $ $ & $ $ & $ $ \\
$\quad {\rm Package} $ & ${\rm Periodic/Basis~set} $ & $\quad {\rm Standard} $ & $\quad {\rm Meta-GGA} $ & $\quad {\rm Hybrid} $ & $\quad {\rm Dispersion-corrected}$ & $\quad {\rm Many-body}$ \\
$ $ & $ $ & $ $ & $ $ & $ $ & $ $ & $ $ \\
\hline
$ $ & $ $ & $ $ & $ $ & $ $ & $ $ & $ $ \\
${\rm ABINIT}$~\cite{abinit}  & $\surd / {\rm PW}$ & $\surd$ & $\surd$ & $\surd$ & $\surd$ & $\surd$ \\
$ $ & $ $ & $ $ & $ $ & $ $ & $ $ & $ $ \\
${\rm ADF}$~\cite{adf}  &    $\surd / {\rm STO}$ & $\surd$ & $\surd$ & $\surd$ & $\surd~(\times~{\rm vdW-DF})$ & $\times$ \\
$ $ & $ $ & $ $ & $ $ & $ $ & $ $ & $ $ \\
${\rm CASTEP}$~\cite{castep}  & $\surd / {\rm PW}$ & $\surd$ & $\surd$ & $\surd$ & $\surd~(\times~{\rm vdW-DF})$ & $\times$ \\
$ $ & $ $ & $ $ & $ $ & $ $ & $ $ & $ $ \\
${\rm CP2K}$~\cite{cp2k}  & $\surd / {\rm GTO,PW}$ & $\surd$ & $\surd$ & $\surd$ & $\surd$ & $\times$ \\
$ $ & $ $ & $ $ & $ $ & $ $ & $ $ & $ $ \\
${\rm CRYSTAL}$~\cite{crystal}  & $\surd / {\rm GTO}$ & $\surd$ & $\surd$ & $\surd$ & $\surd~(\times~{\rm vdW-DF})$ & $\times$ \\
$ $ & $ $ & $ $ & $ $ & $ $ & $ $ & $ $ \\
${\rm GAMESS}$~\cite{gamess}  & $\times / {\rm GTO}$ & $\surd$ & $\surd$ & $\surd$ & $\surd~(\times~{\rm vdW-DF})$ & $\times$ \\
$ $ & $ $ & $ $ & $ $ & $ $ & $ $ & $ $ \\
${\rm GAUSSIAN}$~\cite{gaussian}  & $\surd / {\rm GTO}$ & $\surd$ & $\surd$ & $\surd$ & $\surd~(\times~{\rm vdW-DF})$ & $\times$ \\
$ $ & $ $ & $ $ & $ $ & $ $ & $ $ & $ $ \\
${\rm GPAW}$~\cite{gpaw}  & $\surd / {\rm NAO,PW}$ & $\surd$ & $\surd$ & $\surd$ & $\surd$ & $\surd$ \\
$ $ & $ $ & $ $ & $ $ & $ $ & $ $ & $ $ \\
${\rm MOLPRO}$~\cite{molpro}  & $\times / {\rm GTO}$ & $\surd$ & $\surd$ & $\surd$ & $\times$ & $\times$ \\
$ $ & $ $ & $ $ & $ $ & $ $ & $ $ & $ $ \\
${\rm NWCHEM}$~\cite{nwchem}  & $\surd / {\rm GTO,PW}$ & $\surd$ & $\surd$ & $\surd$ & $\surd~(\times~{\rm vdW-DF})$ & $\times$ \\
$ $ & $ $ & $ $ & $ $ & $ $ & $ $ & $ $ \\
${\rm EXPRESSO}$~\cite{expresso}  & $\surd / {\rm PW}$ & $\surd$ & $\surd$ & $\surd$ & $\surd$ & $\times$ \\
$ $ & $ $ & $ $ & $ $ & $ $ & $ $ & $ $ \\
${\rm SIESTA}$~\cite{siesta}  & $\surd / {\rm NAO}$ & $\surd$ & $\surd$ & $\surd$ & $\surd$ & $\times$ \\
$ $ & $ $ & $ $ & $ $ & $ $ & $ $ & $ $ \\
${\rm VASP}$~\cite{vasp}  & $\surd / {\rm PW}$ & $\surd$ & $\surd$ & $\surd$ & $\surd$ & $\surd$ \\
$ $ & $ $ & $ $ & $ $ & $ $ & $ $ & $ $ \\
${\rm WIEN2K}$~\cite{wien2k}  & $\surd / {\rm FP-(L)APW+lo}$ & $\surd$ & $\surd$ & $\surd$ & $\surd~(\times~{\rm vdW-DF})$ & $\times$ \\
$ $ & $ $ & $ $ & $ $ & $ $ & $ $ & $ $ \\
\hline
\hline
\end{tabular}
\end{center}
\caption{Selected computer packages which can be used to perform DFT calculations. ``Periodic'' refers 
         to the ability of handling the simulation of three-dimensional periodic systems. In ``Basis set'',
         ``PW'' refers to plane waves, ``STO'' to Slater-type orbitals, ``GTO'' to Gaussian orbitals,
          ``NAO'' to numerical atomic orbitals, and ``FP-(L)APW+lo '' to augmented plane waves and local 
         orbitals. Symbol $\surd$ ($\times$) indicates suitability (unsuitableness) of the package to perform 
         a particular calculation type.}
\end{table*}

\section{Assessing the performance of DFT methods in the design of GAM}
\label{sec:assessmentDFT}
Making judicious comparisons between zero-temperature calculations and ambient-temperature 
observations turns out to be very complicate due to the presence of thermal excitations. 
Temperature profoundly affects the (free) energy and conformation of molecules adsorbed on GAM, 
and the differences with respect to $T = 0$ conditions are difficult to assess with theory even 
when considering the simplest interaction models. In fact, free energies and entropies cannot 
be accessed straightforwardly during molecular dynamic simulations, in contrast to other 
thermodynamic quantities like, for instance, the total internal energy and pressure.  
In order to evaluate free energies technically and computationally involved methods like 
thermodynamic integration, free-energy perturbation and umbrella sampling, have to be 
employed~\cite{kastner09,taioli07,cazorla07,cazorla12b}. Reasonably then, the most direct and exact way 
of evaluating the performance of approximate ground-state methods is to compare them with other computational 
approaches which are known to be more accurate. In the case of DFT methods, accuracy benchmark tests 
involve the application of quantum chemistry methods (QCM) like M\o ller-Plesset perturbation theory 
(MP2), the coupled-cluster method with single, double and perturbative triple excitations 
[CCSD(T)], and quantum Monte Carlo (e.g., diffusion Monte Carlo -DMC-). 

In what follows, we review recent DFT benchmark studies done in the areas of hydrogen storage and 
carbon capture and sequestration problems involving CN and MOF. Based on them, we draw general 
conclusions on the suitability of approximate exchange-correlation functionals for the design of GAM. 
Regrettably, despite that the number of these benchmark studies is very modest, as it will be shown next, 
some interpretation and computational inconsistencies prevail yet among them. Also, we discuss
the subtle problems found in the generalization of DFT performance tests done in model 
cluster systems to extended GAM. A few modeling strategies are proposed to overcome this class of 
problems, some of which are computationally very intensive and some others tentative. 

\subsection{H$_{2}$ storage}
\label{subsec:h2-storage-benchmark}

\subsubsection{Carbon-based GAM}
\label{subsec:carbon-gam-h2}
A promising alternative to functionalizing carbon-based GAM with transition  
metals atoms is to use alkali metal species like lithium and calcium. Aggregation issues 
on alkali metal coatings are expected to be less severe than in the transition metal 
case (see Sec.~\ref{subsec:c-b-n}), and the corresponding H$_{2}$ 
gravimetric densities predicted in DFT studies largely surpass the targets set by
the U.S. Department of Energy (i.e., 5.5 wt.~\%)~\cite{ataca09,zhou12,ao10}.  
Reported experimental records, however, lie significantly below the impressive 
performances anticipated with theory (e.g., a modest $2$~wt.~\%)~\cite{chen13,hong12,huang13,zhou14} 
and the reasons for those large discrepancies are not yet totally understood. 

\begin{figure}
\centerline{
\includegraphics[width=1.00\linewidth]{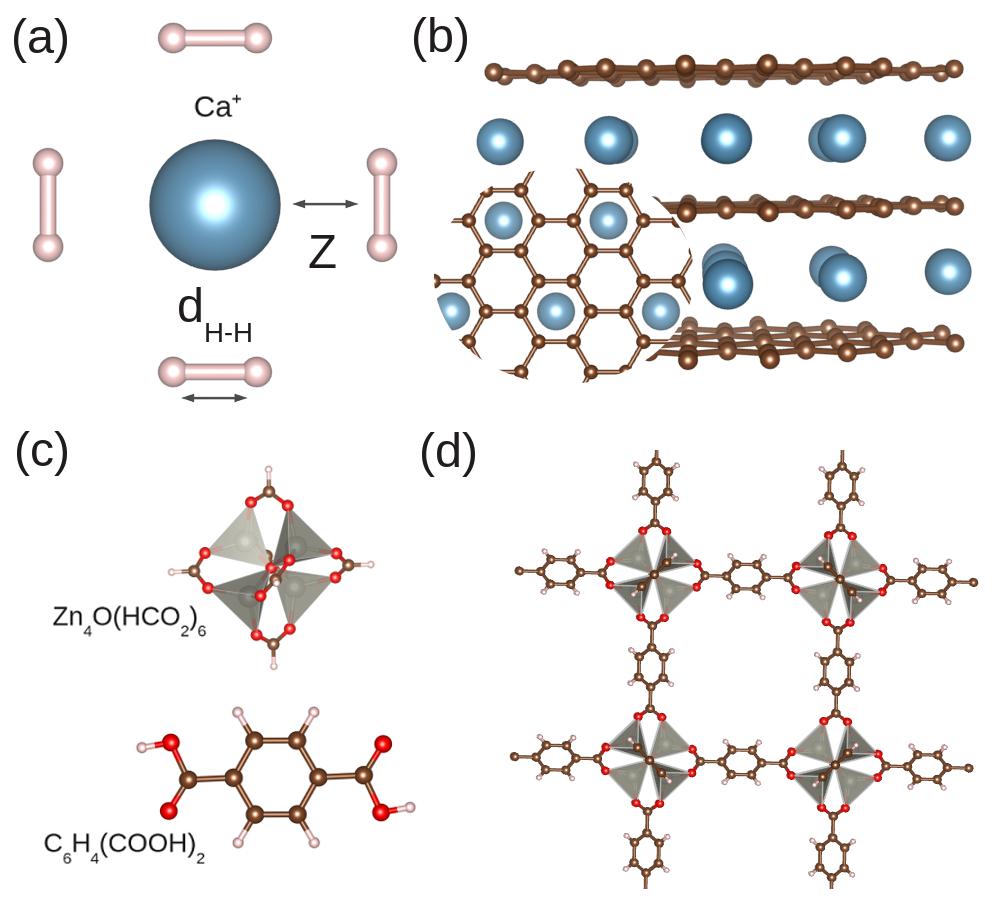}}
\caption{Sketch of the (a)~Ca$^{+}$(H$_{2}$)$_{4}$ model cluster system, (b)~graphite intercalation
         compounds~\cite{dresselhaus02,emery05,cobian08} seen from the front and top views, 
         (c) organic connector and metal cluster model systems, and (d) MOF-5 [also known 
         as IRMOF-1]. Cluster systems in (a) and (c) have been widely considered in benchmark
         studies as representative models of GAM depicted in (b) and (d), respectively.}
\label{cluster-to-material}
\end{figure}

In 2009, Cha \emph{et al.} published a controversial work in which the accuracy of standard 
DFT methods in the assessment of H$_{2}$-storage GAM was seriously questioned~\cite{cha09}.
By using quantum chemistry [i.e., MP2 and CCSD(T)] and standard and hybrid DFT methods, 
Cha showed that the binding energies, $E_{\rm bind}$, and interatomic distances 
calculated in a model system composed of four equidistant hydrogen molecules to a 
positively charged Ca ion [i.e., Ca$^{+}$(H$_{2}$)$_{4}$, see 
Fig.~\ref{cluster-to-material}a], were dramatically  
different. In particular, with standard and hybrid DFT functionals favorable 
molecular binding to the calcium cation (i.e., $E_{\rm bind}$ values around $\sim -1.0$ 
and $-0.1$~eV) was found at a H$_{2}$-Ca$^{+}$ distance of $z = 2.3$~\AA~, 
whereas with MP2 and CCSD(T) methods no effective binding was determined
at any separation (see Table~IV). Cha \emph{et al.} performed analogous  
gas-adsorption calculations in a larger system comprising a coronene molecule 
(i.e., C$_{24}$H$_{12}$) and a calcium atom, and concluded with the same level 
of inconsistency that found in the  Ca$^{+}$(H$_{2}$)$_{4}$ case. In words of 
Cha \emph{et al.}~\cite{cha09}: 
``(these findings) indicate that previous suggestions for the Ca-based hydrogen storage
system should be reinvestigated with particular care about the charge state of Ca''.  
The levels of alarm associated with this error class, however, were lowered shortly 
afterwards by Ohk \emph{et al.}. In a formal comment on Cha's work, Ohk argued that 
the discrepancies between MP2 and DFT methods reported in the Ca$^{+}$(H$_{2}$)$_{4}$ 
and Ca-C$_{24}$H$_{12}$ systems were numerical artifacts stemming from the use of small 
localized orbitals basis sets, i.e., $6-311++G^{**}$, which did not contain polarization 
functions of high enough momenta~\cite{ohk10}. By using larger basis sets of the Dunning type 
and performing extrapolation to the complete basis set (CBS) limit to get rid of 
likely finite basis-set errors, Ohk \emph{et al.} showed that the agreement between  
semilocal DFT and MP2 results obtained in Ca-decorated cluster models was qualitatively 
acceptable (although standard DFT methods exhibited a strong tendency towards molecular
overbinding, see Table~IV). The reliability of GGA functionals in the assessment of hydrogen 
storage materials apparently had it been restored (at least, at the qualitative level). However, a deep 
understanding about how standard DFT functionals, which definitely incur in self-interaction 
errors and neglect long-range dispersive interactions, could describe the adsorption of 
light molecules on surfaces and cavities correctly, was still missing.   

In two subsequent works, the authors of article~\cite{cha09} reported further details on the 
anomalous performance of customary DFT methods in describing the fixation of H$_{2}$ molecules on Ca 
centers via the Kubas interaction~\cite{cha10,cha11}.
They showed that when a sharp orbital transition occurs during adsorption of the gas molecule on the 
metal center, the incompleteness of the electronic exchange, present in most DFT functionals, acts critically 
by providing erroneous overstabilization of the final complex. According to Cha \emph{et al.},  
Ca$^{+}$(H$_{2}$)$_{4}$ exemplifies the type of system where the highest occupied molecular 
orbital (HOMO) changes abruptly from $4s$ to $3d$ character upon the intake of gas, and the fixation of 
H$_{2}$ molecules via the Kubas interaction becomes frustrated. 
Cha's explanations are based on arguments put forward by Gunnarsson more than two decades ago, 
who showed that in situations where the nodal wavefunction surfaces are intricate 
approximate DFT exchange functionals tend to be imprecise by underestimating the 
energy cost associated to orbital transitions~\cite{gunnarsson85}. Cha's arguments indeed  
brought new physical insight into the Ca$^{+}$(H$_{2}$)$_{4}$ contention, however some technical 
aspects of works~\cite{cha10,cha11} could still be criticized (e.g., extrapolation to the CBS limit in 
the MP2 calculations was not pursued). Thus, further benchmark studies were still required for
carefully judging the accuracy of customary DFT methods in the design of carbon-based GAM.     

\begin{table*}
\begin{center}
\label{tab:cah2-4}
\begin{tabular}{c c c c c}
\hline
\hline
$ $ & $ $ & $ $ & $ $ & $ $  \\
$\quad {\rm Work} \quad$ & $\quad E_{\rm bind}~{\rm (eV/H_{2})} \quad$ & $\quad z_{\rm min}$~(\AA) & $\quad d_{\rm H-H}$~(\AA) & $\quad {\rm Method} \quad$  \\
$ $ & $ $ & $ $ & $ $ & $ $  \\
\hline
$ $ & $ $ & $ $ & $ $ & $ $  \\
\cite{cha09}& $-1.50$ & $2.3$ & $0.81$ & ${\rm LDA/6-311++G**}$     \\
$ $       & $-0.30$ & $2.3$ & $0.78$ & ${\rm PBE/6-311++G**}$     \\
$ $       & $-0.40$ & $2.3$ & $0.77$ & ${\rm B3LYP/6-311++G**}$   \\
$ $       & $-0.05$ & $4.2$ & $0.74$ & ${\rm MP2/6-311++G**}$     \\
$ $       & $-0.05$ & $4.2$ & $0.74$ & ${\rm CCSD(T)/6-311++G**}$ \\
$ $ & $ $ & $ $ & $ $ & $ $  \\
\hline
$ $ & $ $ & $ $ & $ $ & $ $  \\
\cite{ohk10}& $-0.90$ & $2.3$ & $ $ & ${\rm PBE/cc-pVQZ}$  \\
$ $       & $-0.20$ & $3.4$ & $ $ & ${\rm MP2/cc-pVQZ}$  \\
$ $       & $-0.15$ & $2.3$ & $ $ & ${\rm CCSD(T)/CBS}$  \\
$ $ & $ $ & $ $ & $ $ & $ $  \\
\hline
$ $ & $ $ & $ $ & $ $ & $ $  \\
\cite{bajdich10}& $-1.20$ & $2.3$ &       $0.77$ & ${\rm LDA}$      \\
$ $       & $-0.70$ & $2.3$ &       $0.77$ & ${\rm PBE}$      \\
$ $       & $-0.30$ & $2.3$ &       $0.77$ & ${\rm B3LYP}$    \\
$ $       & ${\rm No~binding} $ & $z \le 4.6$ & $0.77$ & ${\rm MP2}$      \\
$ $       & ${\rm No~binding} $ & $z \le 4.6$ & $0.77$ & ${\rm HFx-PBEc}$ \\
$ $       & ${\rm No~binding} $ & $z \le 4.6$ & $0.77$ & ${\rm DMC/ANO-VTZ}$      \\
$ $ & $ $ & $ $ & $ $ & $ $  \\
\hline
$ $ & $ $ & $ $ & $ $ & $ $  \\
\cite{purwanto11}& $-0.16$ & $3.5$ & $0.74$ & ${\rm MP2/CBS}   $  \\
$ $       & $-0.18$ & $2.2/3.4$ & $0.77$ & ${\rm AF-QMC/CBS}$  \\
$ $ & $ $ & $ $ & $ $ & $ $  \\
\hline
\hline
\end{tabular}
\end{center}
\caption{Summary of the binding energy and structural results obtained in the 
Ca$^{+}$(H$_{2}$)$_{4}$ model cluster system by different authors employing a variety of 
DFT and quantum chemistry methods (see text).}
\end{table*}

In this regard, Bajdich \emph{et al.} made an important contribution by performing 
for the first time quantum Monte Carlo (QMC) calculations in the Ca$^{+}$(H$_{2}$)$_{4}$
model system, and comparing their results to those obtained with the MP2, and local, 
semilocal and hybrid DFT methods~\cite{bajdich10} (see Table~IV). It was found that QMC 
calculations based on the diffusion Monte Carlo (DMC) predicted no binding at all of the four 
H$_{2}$ molecules to the Ca$^{+}$ center within the interval $0 \le z \le 4.6$~\AA~, 
in agreement with previous MP2 results. In stark contrast to this conclusion, popular DFT functionals 
like LDA, GGA and B3LYP vaticinated effective fixation of the gas molecules at a distance of $2.3$~\AA~. 
Bajdich's results are in qualitative agreement~(disagreement) with Cha's~(Ohk's) conclusions  
explained above. Interestingly, the authors of work~\cite{bajdich10} carried out DFT calculations  
with a blended functional consisting of the full HF exchange energy and the PBE correlation functional
(denoted as HFx-PBEc in Table~IV). They showed that the results obtained with the HFx-PBEc functional were 
consistent with those obtained with the MP2 and DMC methods (see Table~IV). In light of those outcomes, 
Bajdich \emph{et al.} argued that the failure of common DFT functionals in describing the Ca$^{+}$(H$_{2}$)$_{4}$ 
system had its origins on the partial or total omission of long-range exchange interactions, in coincidence
with Cha's reasonings found in works~\cite{cha10,cha11}. 
In spite of the significance of Bajdich's work, this is neither free of some technical objections. 
For instance, in the atomic relaxations the H-H intermolecular distances were kept fixed to $0.77$~\AA, 
and the errors stemming from the incompleteness of the employed basis sets (i.e., triple-zeta) and the 
fixed-node surface approximation in DMC, were not evaluated. 

In a posterior work, Purwanto \emph{et al.} presented a highly accurate study on the binding of the 
Ca$^{+}$(H$_{2}$)$_{4}$ complex which relied on auxiliary-field QMC calculations~\cite{purwanto11}.   
There, extrapolation to the complete basis set limit and a better treatment of the sign problem
were accomplished. Purwanto \emph{et al.} found that the potential energy curve of the four hydrogen
molecules exhibits a double-well structure with almost equal binding minima of $\sim -0.18$~eV   
at distances $2.2$ and $3.4$~\AA. These results are in good agreement with the MP2/CBS calculations
performed by the same authors and the CCSD(T)/CBS results obtained by Ohk \emph{et al.}~\cite{ohk10}
(see Table~IV). Regarding the differences with respect to Bajdich's work, it was argued that
these were likely to be originated by the fixed-node approximation employed in the DMC calculations. 

\emph{Conclusions.-}~The main conclusion emerging from works~\cite{cha09,ohk10,cha10,cha11,bajdich10,purwanto11} 
is that standard DFT functionals tend to overestimate the cohesion of the Ca$^{+}$(H$_{2}$)$_{4}$
complex, and that the likely reason for this flaw is a deficient treatment of the long-range  
exchange energy. On the other hand, highly accurate QCM results obtained by different groups on   
the same system are not fully consistent and therefore a conclusive verdict on the general 
performance of DFT methods, i.e., considering all its possible flavours, cannot be emitted. The current 
status of computational work on this topic is clearly unsatisfactory.

Yet, the situation becomes even more puzzling when the resemblance between the model system
Ca$^{+}$(H$_{2}$)$_{4}$ and real carbon-based H$_{2}$-storage GAM is brought into examination (e.g., see 
Fig.~\ref{cluster-to-material}a-b). Namely, there is still the unresolved question about on which 
grounds the results obtained in model cluster systems can be translated (if possible at all) to realistic
extended materials~\cite{ma11,smith13}. For instance, the likely presence of valence $s$ and $p$ 
electronic states, coming from the carbon atoms in the sorbent, is totally disregarded in the 
Ca$^{+}$(H$_{2}$)$_{4}$ complex. Also, the role of the long-range dispersion interactions, which are 
ubiquitous in gas-adsorption processes, normally turns out to be underestimated in nano-sized systems. 
In this last regard, several analysis on the hydrogen storage properties of extended Ca-decorated 
carbon nanomaterials based on dispersion-corrected DFT functionals, have been reported recently. 

Wang \emph{et al.} have studied the H$_{2}$-storage properties of Ca-decorated graphene with the vdW-DF 
approach due to Dion~\cite{dion04} and compared their results to those obtained with standard local and 
semilocal methods~\cite{wang12}. Wang's calculations show that the H$_{2}$ binding energies obtained 
with the vdW-DF method lie systematically below those obtained with GGA-PBE (i.e., stronger molecule-GAM 
interactions by $\sim 0.05$~eV/molecule), and generally above the LDA results (i.e., weaker molecule-GAM 
interactions by $\sim 0.05-0.10$~eV/molecule). (Similar conclusions have been attained by other authors 
in alike Ca-decorated carbon nanomaterials~\cite{hussain12,li11b}.)
In light of these outcomes, it is argued that long-range dispersion interactions can be also 
important in the adsorption of hydrogen molecules on chemically modified carbon nanomaterials, even
in the cases where the obtained binding energies are large (i.e., $|E_{\rm bind}| \ge 0.1$~eV/molecule). 
Actually, Wang's results appear to expose a new failure of standard DFT methods in the assessment
of GAM, this time related to the description of the electronic correlation energy. Unfortunately, the authors 
of work~\cite{wang12} did not report any comparison with respect to hybrid functionals or QCM calculations, 
hence the size of the bias incurred by the vdW-DF method itself, i.e., due to the approximations performed on 
the exchange functional and pairwise additivity, cannot be inferred. 

In work~\cite{wong14}, Wong \emph{et al.} have investigated also the adsorption of H$_{2}$ molecules 
on metal-decorated graphene but considering a large set of transition and alkali metal species. 
The employed method there is an improved version of Dion's approach, the so-called vdW-DF2, in which the 
accuracy of the employed semilocal exchange functional is bettered~\cite{lee10b}. In the case of Ca-based 
coatings, Wong's results are in qualitative agreement with those obtained by Wang \emph{et al.}~\cite{wang12} 
however the difference between the two reported vdW-DF and vdW-DF2 H$_{2}$-binding energies is not negligible 
(i.e., about $30$~meV). When considering other types of dopants, vdW-DF2 calculations generally predict 
weaker molecular binding than obtained with semilocal GGA methods by $\sim 0.1$~eV per molecule. 
In fact, Wong's work comes to reinforce the idea that dispersion interactions can affect profoundly 
the interplay between extended materials and hydrogen molecules. Consequently, nonlocal correlations 
must be taken into account by any DFT functional that is intended for modeling of GAM. 

\emph{Conclusions.-}~From all these benchmark studies, we can draw the general conclusion that both long-range 
exchange and dispersion electronic interactions are pivotal in describing the binding of H$_{2}$ molecules to CN-based 
GAM. As for local and semilocal DFT functionals, these two elements are totally missing in them, therefore they 
are likely to provide unreliable results on the hydrogen storage topic. Situations in which local and semilocal GAM predictions  
seem to be correct normally are fortuitous and indicate the presence of large energy error cancellations~\cite{hussain12,li11b}. 
Namely, the sign of the missing contributions to the exchange-correlation energy in local and semilocal approaches are 
opposite and therefore can compensate each to the other. These error cancellations, however, do not occur systematically 
and may depend on the specific details of GAM (e.g., see work~\cite{wong14} where the sign of the difference between 
the GGA and vdW-DF2 energies varies with the dopant species), and thus the use of standard DFT methods for modeling 
of carbon-based H$_{2}$-storage materials is not recommended. Meanwhile, new quantum Monte Carlo (QMC) and RPA-DFT 
simulations, this time performed in periodic systems, are highly desirable for rigorously evaluating the performance
of meta, hybrid, and dispersion DFT functionals (see works~\cite{schimka10,scheffer12,wu09,binnie10,wagner14} for examples 
of applications of such advanced computational methods to simulation of relevant materials). QMC and RPA-DFT calculations 
are also necessary for determining the relevance of many-body energy and Coulomb screening effects on the present 
class of GAM, which so far have been systematically neglected. 
 
\subsubsection{MOF}
\label{subsec:mof-h2}
The interplay between H$_{2}$ molecules and MOF (e.g., MOF-74,5) are dominated by dispersion 
interactions and the local environment surrounding the binding sites in the metal 
clusters~\cite{kuc08a,kuc08b,sillar09,joo13}. Consequently, employing DFT methods that completely 
neglect nonlocal correlations (i.e., standard and hybrid flavours) and/or carrying out benchmark 
tests in model cluster systems which are too small (see Fig.~\ref{cluster-to-material}c-d) turns 
out to be inadequate in the present case. On the other side, the fact that the H$_{2}$-MOF interactions 
are weak implies that effective gas-storage can only be achieved at temperatures well below ambient 
conditions (i.e., $T \sim 80$~K). Thus, a number of strategies have been proposed for increasing the 
affinity of MOF towards hydrogen binding~\cite{rowsell05}. These include, the design and control of 
porosity~\cite{han09}, functionalization of the organic linkers~\cite{han08}, hydrogen 
spillover~\cite{li07,wang12b}, introduction of open metal sites in the organic linkers and metal
clusters~\cite{kaye08,lochan08,zhou08,sun07,kumar12}, and decoration of MOF surfaces with alkali 
metal atoms~\cite{han07,klontzas08,mulfort08}. DFT methods have been intensively employed in the 
last two mentioned approaches hence we revise next the benchmark tests undertaken in those areas.  

\begin{table*}
\begin{center}
\label{tab:h2-mof}
\begin{tabular}{c c c c c c c c c}
\hline
\hline
$ $ & $ $ & $ $ & $ $ & $ $ & $ $ & $ $ & $ $ & $ $ \\
${\rm Work}$ & ${\rm System}$ & \multicolumn{7}{c}{$E_{\rm bind}~{\rm (eV/H_{2})}$} \\
$ $ & $ $ & $ $ & $ $ & $ $ & $ $ & $ $ & $ $ & $ $ \\
$ $ & $ $ & ${\rm PW91}$ & ${\rm PBE}$ & ${\rm B3LYP}$ & ${\rm M05-2X}$ & ${\rm MP2}$ & ${\rm CCSD(T)}$ & $\quad {\rm DMC} \quad$ \\
$ $ & $ $ & $ $ & $ $ & $ $ & $ $ & $ $ & $ $ & $ $ \\
\hline
$ $ & $ $ & $ $ & $ $ & $ $ & $ $ & $ $ & $ $ & $ $ \\
\cite{sun09}   & ${\rm H_{2}-Ca@H_{4}B_{2}C_{6}O_{4}}$ & $ $ & $-0.16$ & $ $ & $ $ & $-0.16$ & $ $ & $ $ \\
$ $ & $ $ & $ $ & $ $ & $ $ & $ $ & ${\rm (CBS)} $ & $ $ & $ $ \\
$ $ & $ $ & $ $ & $ $ & $ $ & $ $ & $ $ & $ $ & $ $ \\
\cite{sun10}   & ${\rm H_{2}-Ca@C_{8}H_{6}O_{4}}$      & $-0.22$ & $-0.19$ & $-0.15$ & $-0.26$ & $-0.26$ & $-0.24$ & $ $ \\
$ $ & $ $ & $ $ & $ $ & $ $ & $ $ & ${\rm (CBS)} $ & ${\rm (CBS)} $ & $ $ \\
$ $ & $ $ & $ $ & $ $ & $ $ & $ $ & $ $ & $ $ & $ $ \\
\cite{sun10}   & ${\rm H_{2}-Sc@C_{5}H_{5}}$           & $-0.28$ & $-0.27$ & $-0.14$ & $-0.23$ & $-0.24$ & $-0.23$ & $ $ \\
$ $ & $ $ & $ $ & $ $ & $ $ & $ $ & ${\rm (CBS)} $ & ${\rm (CBS)} $ & $ $ \\
$ $ & $ $ & $ $ & $ $ & $ $ & $ $ & $ $ & $ $ & $ $ \\
\cite{sun10}   & ${\rm H_{2}-Ti@C_{2}H_{4}}$           & $-0.39$ & $-0.37$ & $-0.26$ & $-0.38$ & $-0.42$ & $-0.37$ & $ $ \\
$ $ & $ $ & $ $ & $ $ & $ $ & $ $ & ${\rm (CBS)} $ & ${\rm (CBS)} $ & $ $ \\
$ $ & $ $ & $ $ & $ $ & $ $ & $ $ & $ $ & $ $ & $ $ \\
\cite{sun10}   & ${\rm H_{2}-Li@C_{8}H_{6}O_{4}}$      & $-0.15$ & $-0.13$ & $-0.10$ & $-0.16$ & $-0.16$ & $-0.16$ & $ $ \\
$ $ & $ $ & $ $ & $ $ & $ $ & $ $ & ${\rm (CBS)} $ & ${\rm (CBS)} $ & $ $ \\
$ $ & $ $ & $ $ & $ $ & $ $ & $ $ & $ $ & $ $ & $ $ \\
\cite{dixit11} & ${\rm H_{2}-Li@C_{10}H_{10}O_{4}}$    & $-0.42$ & $-0.18$ & $-0.16$ & $ $ & $-0.17$ & $ $ & $ $ \\
$ $ & $ $ & ${\rm (cc-pVTZ)} $ & $ $ & ${\rm (cc-pVTZ)} $ & $ $ & ${\rm (cc-pVTZ)} $ & $ $ & $ $ \\
$ $ & $ $ & $ $ & $ $ & $ $ & $ $ & $ $ & $ $ & $ $ \\
\cite{jiang12} & ${\rm H_{2}-Li@C_{4}H_{3}}$           & $-0.11$ & $-0.12$ & $-0.12$ & $ $ & $ $ & $ $ & $-0.14$ \\
$ $ & $ $ & $ $ & $ $ & ${\rm (6-311G[d,p])} $ & $ $ & $ $ & $ $ & $ $ \\
$ $ & $ $ & $ $ & $ $ & $ $ & $ $ & $ $ & $ $ & $ $ \\
\hline
\hline
\end{tabular}
\end{center}
\caption{Results of different benchmark tests carried out in organic linker MOF models decorated with  
alkaline earth, alkali and transition metal atomic species. The types of localized-orbitals basis sets
employed in the calculations and the cases in which convergence to the CBS limit is achieved, 
are indicated within parentheses.} 
\end{table*}
        
It was first argued by Lochan \emph{et al.} that H$_{2}$ molecules are attracted by open transition 
metal and alkali centers in MOF through donor-acceptor interactions and electrostatics~\cite{lochan08}.  
Based on the outcomes of standard DFT calculations, those authors concluded that the strength of the  
H$_{2}$-MOF interactions was within the range of desirable binding for ambient gas storage applications 
(i.e., $0.3-0.8$~eV). Lochan \emph{et al.} also claimed that semilocal DFT functionals could be \emph{safely} 
employed in the study of H$_{2}$-MOF systems because the corresponding leading electronic interactions 
are strong and predominant over dispersion~\cite{lochan08}.  

In two recent studies, Sun \emph{et al.} have evaluated the accuracy of standard, hybrid and meta DFT 
functionals in the prediction of H$_{2}$ adsorption on metal-doped organic linker systems~\cite{sun09,sun10}. 
They have found that whenever transition metal, alkaline-earth or alkali metal atoms are used,  
all DFT, MP2 and CCSD(T) methods provide quantitatively similar binding energy results, 
in satisfactory accordance with Lochan's findings (see Table~V). In particular, calculations 
performed with popular DFT functionals like PBE, PW91 and M05-2X are in excellent agreement with gold-standard 
benchmarks obtained with the CCSD(T) method (i.e., equal binding energies to within $\sim 0.01$~eV). Meanwhile, hybrid 
functionals tend to underestimate $E_{\rm bind}$ slightly by $0.05-0.10$~eV. Analogous conclusions have been 
attained also by Dixit \emph{et al.} in a posterior work done in Li-decorated MOF (see Table~V)~\cite{dixit11}.  

A further benchmark test confirming the accuracy of DFT methods in describing the interplay between H$_{2}$ molecules
and chemically functionalized organic linkers, has been recently reported by Jiang \emph{et al.}~\cite{jiang12}. 
In particular, Jiang \emph{et al.} have studied the binding of a hydrogen molecule to a small C$_{4}$H$_{3}$Li 
cluster using common GGA and hybrid DFT functionals, and the highly accurate DMC method. As it can be appreciated 
in Table~V, notable agreement between all the considered approaches in Jiang's calculations is obtained (i.e., 
within $\sim 0.01$~eV).  

\emph{Conclusions.-}~In light of the benchmark results reported in works~\cite{sun09,sun10,dixit11,jiang12}, we may 
conclude that customary DFT methods appear to perform appropriately in the simulation of the hydrogen-storage properties 
of chemically functionalized MOF. The physical reason underlying this favorable outcome is that donor-acceptor 
interactions and electrostatics in this family of GAM are dominated by short- and medium-range 
electron-electron exchange and correlations. Nevertheless, a note of caution must be added here. 

In our compilation of benchmark tests we have realized a lack of studies analyzing the performance of dispersion-corrected 
DFT schemes in the simulation of extended hydrogen-loaded MOF containing open metal sites. Indeed, dispersion interactions 
appear to be secondary in the present case however, as we will show in Sec.~\ref{subsec:mof-ccs}, when H$_{2}$ molecules 
are replaced by CO$_{2}$ this type of interactions turns out to be crucial. The polarizability of the H$_{2}$ 
molecule certainly is smaller than that of CO$_{2}$ ($0.79$ and $2.51$~\AA$^{3}$, respectively~\cite{olney97}), 
however this still must have some effect. Actually, it would be very interesting to quantify in which proportion 
dispersion interactions tend to lower the DFT energies reported in Table~V (which, we remind, have been obtained in model 
cluster systems). Also, it would be highly desirable to perform systematic studies on the performance of local, semilocal, 
and nonlocal DFT approaches in periodic simulation of chemically functionalized MOF loaded with hydrogen. 
In this regard, we would like to mention a recent work by Sumida \emph{et al.} in which it has been shown that 
neither standard nor hybrid DFT functionals can reproduce with accuracy the measurements done on the binding 
of H$_{2}$ molecules to metal-BTT~\cite{sumida13}. Rather, a range-separated hybrid and dispersion corrected 
DFT functional, i.e., the so-called $\omega$B97X-D (see Sec.~\ref{subsec:hybrid}), is found to be necessary for 
a correct interpretation of the experimental findings (see for instance Table~II in work~\cite{sumida13}). Also, 
Kong \emph{et al.} have recently found that nonlocal interactions are crucial to achieve DFT consistency with 
respect to the H$_{2}$ heats of adsorption measured in Zn$_{2}$(BDC)$_{2}$(C$_{6}$H$_{12}$N$_{2}$)~\cite{kong09}. 
Further DFT work on the role of the dispersion interactions in hydrogen storage of chemically functionalized MOF, 
is urgently needed in order to avoid likely modeling inconsistencies.

\subsection{CO$_{2}$ capture and sequestration}
\label{subsec:ccs-benchmark}

\subsubsection{Carbon-based GAM}
\label{subsec:carbon-gam-ccs}
The adsorption of CO$_{2}$ molecules on carbon-based GAM can be of physisorption or chemisorption type
depending on whether the material surfaces are smooth, contain defects or are chemically functionalized.
In the case of nondefective surfaces, the interactions with the gas molecules are dominated by dispersion
forces thereby the gas is retained on the adsorbent material very weakly~\cite{mishra11a}. 

In 2006, Xu \emph{et al.} were the first in carrying out nonstandard DFT calculations on the binding of 
CO$_{2}$ molecules to pristine graphene~\cite{xu06}. By using the hybrid ONIOM[B3LYP:DFTB-D] method, they 
found that the corresponding gas-adsorption energy was $E_{\rm bind} = -0.03$~eV. In Xu's approach, an hybrid 
DFT evaluation of the interactions between CO$_{2}$ and a coronene molecule is first performed, and 
a tight-binding dispersion correction is subsequently added in order to account for the presence of 
$\pi$-conjugated interactions in the real material. More recently, Umadevi \emph{et al.} have analyzed the 
same type of problem but employing meta-GGA DFT methods (i.e., M05-2X), and surprisingly they have found a 
large physisorption energy of $\sim -0.10$~eV~\cite{umadevi11}.
The reasons for the three-fold discrepancy between Xu's and Umadevi's results remain unclear to us
since QCM benchmark calculations on the strength and nature of the CO$_{2}$-graphene or CO$_{2}$-coronene
interactions are practically absent to date. 
We only know of a recent work~\cite{lee13} by Lee \emph{et al.} in which the physisorption energy of 
CO$_{2}$ is calculated with the MP2 method. In this case, $E_{\rm bind}$ turns out to amount ot $-0.09$ 
and $-0.13$~eV, depending on the relative orientation between the gas molecule and carbon plane. Nevertheless, 
the basis set of localized orbitals employed in Lee's study is relatively small (i.e., 6-31G**) and the model 
cluster system in which the MP2 calculations are performed contains carbon dangling bonds (in opposition to 
real graphene).

On the other side, the adsorption of gas molecules on carbon nanotubes (CNT) has been thoroughly analyzed
by Qui\~{n}onero \emph{et al.} with dispersion-corrected DFT methods (i.e., B97-D/SVP)~\cite{quinonero12}.
By considering different types of CNT, diameters, and binding sites, Qui\~{n}onero \emph{et al.} have
concluded that CO$_{2}$-adsorption is energetically more favourable in the interior than in the exterior
of nanotubes, in marked disagreement with previous reports~\cite{bienfait04,zhao02}. Also, they have found
that the strongest CO$_{2}$-CNT interactions occur in the $(9,0)$ and $(5,5)$ systems where the computed 
binding energy amounts to $\sim -0.6$~eV. Interestingly, based on a symmetry-adapted perturbation theory (SAPT) 
decomposition of the calculated DFT-D interaction energies, the authors of work~\cite{quinonero12} show that 
(i)~dispersion interactions account for the $80$~\% of the total attractive forces, and (ii)~electrostatic 
interactions resulting from the overlap between CO$_{2}$ and CNT electronic orbitals, although secondary, are
yet important. We realize, however, a lack of high-level \emph{ab initio} investigations, e.g., employing the 
MP2 and CCSD(T) methods, in the effects of the CNT size and curvature on the binding of CO$_{2}$ molecules.
This type of studies would be highly desirable for a consistent evaluation of the performance of DFT methods 
on this topic. Similar investigations to these proposed have already been performed for methane~\cite{smith13b}, 
a molecule that is akin to carbon dioxide in terms of the electric dipole and quadrupole moments. 
It is worth noticing that during the preparation of this review we became aware of the submission of an 
article by Smith and Patkowski which could fill the mentioned knowledge gap in CO$_{2}$-CNT systems~\cite{smith14}.

When the surfaces of the carbon nanomaterials contain reactive defect sites like vacancies, holes
(i.e., clusters of vacancies) and edges, the interactions with CO$_{2}$ molecules become
more intense. For instance, Cabrera-Sanfelix has studied the adsorption of carbon dioxide on a defected
graphene sheet with standard DFT methods (i.e., DFT-GGA) and has found that molecular physisorption and
chemisorption have an energy cost of $-0.14$ and $-1.44$~eV, respectively~\cite{cabrera09}.
Similar results have been obtained by Liu \emph{et al.} in an equivalent system using an
analogous computational approach (i.e., $-0.21$ and $-1.72$~eV, respectively)~\cite{liu11}. However,
the authors of this last study propose an equilibrium chemisorption configuration that is more symmetric
than the one obtained by Cabrera-Sanfelix. 
In a more recent and technically exhaustive work, Wood \emph{et al.} have addressed the same kind of gas-adsorption 
problem by employing both dispersion corrected DFT and MP2 methods~\cite{wood12}. Essentially, they find 
that DFT methods incorporating van der Waals forces provide adsorption energies which are in good 
agreement with MP2 results, and that CO$_{2}$ physisorption in the edges of a zigzag graphene nanoribbon  
occurs with an energy balance of $-0.20$~eV. 

\emph{Conclusions.-}~In light of the results reported by Cabrera-Sanfelix, Liu, and Wood, we can 
conclude that using standard DFT methods for representing the interactions of CO$_{2}$ molecules 
with defected graphene seems to be appropriate. It is not clear to us, however, whether the suitability of DFT-GGA 
methods in this case corresponds to a large cancellation between errors or simply to a minor role played by the 
electronic long-range exchange and correlations.

Meanwhile, some functionalization techniques have been proposed for increasing the affinity
of carbon-based GAM towards CO$_{2}$ binding, which is desirable for precombustion applications.
Among those we highlight the decoration of carbon surfaces with nitrogen and light metal
atoms, for which a number of experimental and first-principles computational works have been
performed~\cite{du09,chen14,mishra11,cazorla11,yang10b,yong01,gao11,jiao14}.
Mo \emph{et al.} have recently presented a computational benchmark study of the
adsorption of carbon dioxide on nitrogen-containing hydrocarbon molecules~\cite{mo13}.
Specifically, they have carried out extensive CCSD(T)/CBS and dispersion-corrected DFT
calculations in a CO$_{2}$/2-methylpyridine (the last with formula C$_{6}$H$_{7}$N) system.
Stabilization of this complex occurs through an electron donor (2-methylpyridine)-electron
acceptor (CO$_{2}$) mechanism and the attractive forces between oxygen and hydrogen
atoms. The authors of work~\cite{mo13} show that in order to reproduce the $E_{\rm bind}$
gold-standard obtained with the CCSD(T)/CBS method (i.e., $-0.14$~eV), both electronic long-range  
exchange and correlations must be taken into account simultaneously. According to Mo's calculations, 
BLYP-D3 is among the PBE-D3, BP86-D3, and TPSS-D3 methods (see Sec.~\ref{subsec:vdw}) the 
one peforming the best (i.e., providing a binding energy of $-0.13$~eV). 

\emph{Conclusions.-}~The CO$_{2}$/2-methylpyridine complex is therefore a representative example of a
system in which, despite of its reduced length, both van der Waals interactions and electronic long-range 
exchange are simultaneously important. Also, we note that ``low-cost'' dispersion-corrected DFT schemes like 
the BLYP-D3 one~\cite{grimme-d3} appear to work remarkably well in this system. Indeed, it would be extremely 
interesting to check whether the adsorption of CO$_{2}$ (and H$_{2}$) molecules in alike cluster systems could 
be described correctly also with this last type of computationally cost-effective approaches.

\begin{figure}[h]
\centerline{
\includegraphics[width=1.00\linewidth]{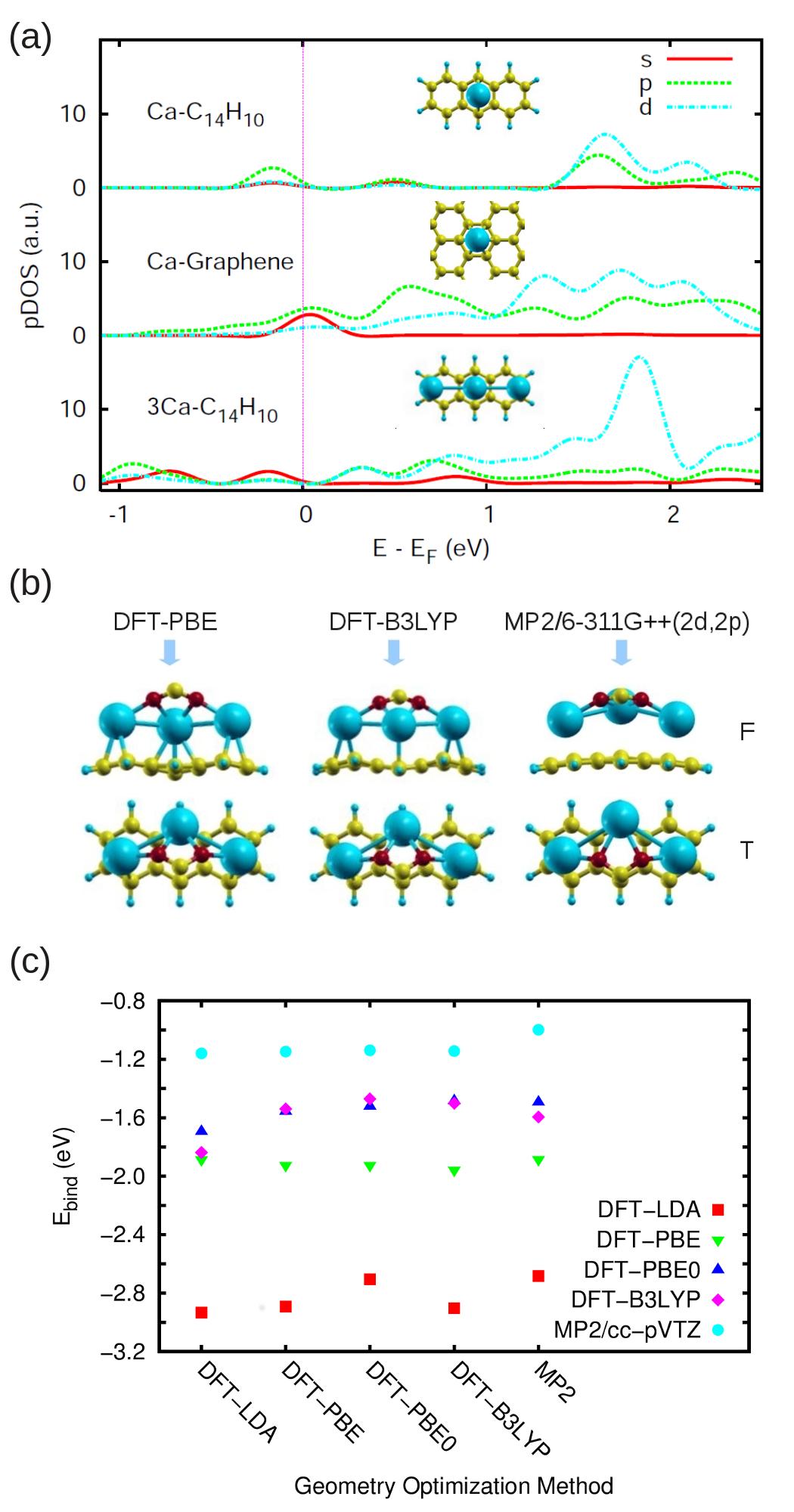}}
\caption{(a)~Partial density of electronic states calculated in Ca-anthracene,
         3Ca-anthracene, and Ca-decorated graphene using DFT-PBE. (b)~Equilibrium 
         structures obtained for the adsorption of CO$_{2}$ in 3Ca-anthracene using 
         DFT and MP2 methods. (c)~Adsorption energies for CO$_{2}$ in 3Ca-anthracene 
         obtained with different geometry optimization and total energy methods.
         (Figure adapted from work~\cite{cazorla13}).}
\label{co2-cns}
\end{figure}

The accuracy of standard and hybrid DFT functionals in describing the interactions of
CO$_{2}$ molecules with Ca-decorated graphene has been recently assessed by Cazorla
\emph{et al.}~\cite{cazorla13}. In Cazorla's work, a comparative study between DFT and MP2 calculations is
presented by following an original recipe: instead of adopting the customary strategy of steadily increasing
the size of polyciclic aromatic hydrocarbon (PAH) molecules, the concentration of Ca dopants in anthracene 
(i.e., C$_{14}$H$_{10}$, a relatively small PAH) is tuned so as to mimic the partial density of electronic 
valence states in Ca-decorated graphene (see Fig.~\ref{co2-cns}a). The reason for doing that is to constrain  
the size of the cluster system where to carry out the MP2 calculations as much as possible, while still 
reproducing the main electronic orbital mechanisms occurring in the targeted extended material. In this way, 
the appearance of artificial electronic transitions upon gas-loading are prevented in the model cluster system 
(e.g., see work~\cite{cha11}). Cazorla's results show that all considered DFT flavors predict equilibrium 
structures which are very similar to that obtained with the MP2 method, and energetically favourable CO$_{2}$ 
binding (see Figs.~\ref{co2-cns}b-c). Nevertheless, the differences between hybrid DFT and MP2 energy estimations 
amount to $\sim 0.4$~eV, and between standard DFT and MP2 to $1.0-2.0$~eV, with DFT methods providing always 
the strongest binding. The origins of the observed hybrid DFT and MP2 numerical discrepancies (see 
Fig.~\ref{co2-cns}c) are rationalized in terms of residual self-interaction errors~\cite{cazorla13}.
The amount of charge transferred between the gas molecules and Ca-C$_{14}$H$_{10}$ turns out to be the same 
when computed with either hybrid DFT or MP2 methods, and long-range dispersion interactions appear to be secondary 
in the present system (see Table~$1$ in work~\cite{cazorla13}). Cazorla \emph{et al.} therefore conclude that 
the strength of the resulting electrostatic interactions, which are dominant, must be equal in the two compared 
cases. On the other side, LDA and PBE standard functionals overestimate the transfer of charge to the CO$_{2}$ 
molecule by $30-40$\%~\cite{cazorla13}. In light of those outcomes, the use of hybrid DFT functionals is 
recommended over that of local and semilocal approaches for investigation of the gas-uptake properties of 
AEM-decorated carbon GAM. 

\emph{Conclusions.-}~Cazorla's prescription for choosing a reduced cluster system in which to 
undertake quantum chemistry calculations may be useful for optimizing the computational expense
associated to DFT benchmark tests. Also, for justifying the subsequent generalization of the
attained conclusions to extended materials~\cite{cazorla13}. Nevertheless, only few selected electronic 
features of the targeted system (e.g., partial density of electronic states around the Fermi energy level) 
can be reproduced at a time by playing with doping strategies, and probably only at a qualitative level. 
In this context, recent methodological progress achieved in the resolution of the so-called ``inverse band 
structure-problem of finding an atomic configuration with given electronic properties'' problem~\cite{franceschetti99,dudly06}
(e.g., genetic algorithm searches), could be useful. In particular, those techniques could be applied to 
the design of model cluster systems that were replicas of the targeted extended GAM in terms of the 
electronic structure. In that case, besides the chemical stoichiometry, the atomic structure and shape of 
the simulation cell could be varied until finding a suitable system in which to perform all the 
calculations. (Of course, electronic long-range exchange and correlation corrections should be added 
somehow afterwards; this could be achieved, for instance, by carrying out additional calculations in 
both periodic and cluster systems with an adequate first-principles method.) We speculate that with such an 
original effective approach it could be possible to address important benchmark controversies affecting the 
performance of DFT methods in GAM modeling (e.g., those explained in the present and previous sections), while 
avoiding the computational burden and bias introduced by the finite size of model cluster systems.

\begin{table*}
\begin{center}
\label{tab:co2-mof}
\begin{tabular}{c c c c c c}
\hline
\hline
$ $ & $ $ & $ $ & $ $ & $ $ & $ $\\
$\quad {\rm Work} \quad$ & $\quad {\rm MOF-type} \quad$ & $\quad {\rm DFT~flavour} \quad$ & $\quad E_{\rm bind} \quad$ & $\quad \Delta H \quad$  & $\quad \Delta H^{\rm exp} \quad$ \\
$ $ & $ $ & $ $ & $ $ & $ $ & $ $\\
\hline
$ $ & $ $ & $ $ & $ $ & $ $ & $ $ \\
\cite{poloni12} & ${\rm Mg-MOF74}$ & ${\rm PBE}$ & $-0.230 $ & $-0.189 $ & $ $ \\
$ $ & $ $ & ${\rm PBE+D2} $                      & $-0.439 $ & $-0.399 $ & $ $ \\
$ $ & $ $ & ${\rm vdW-DF} $                     & $-0.460 $ & $-0.420 $ & $ $ \\
$ $ & $ $ & ${\rm vdW-DF2} $                    & $-0.428 $ & $-0.388 $ & $ $ \\
$ $ & $ $ & ${\rm vdW-PBE} $                     & $-0.644 $ & $-0.604 $ & $ $ \\
$ $ & $ $ & ${\rm vdW-optB88} $                  & $-0.557 $ & $-0.517 $ & $ $ \\
$ $ & $ $ & ${\rm vdW-C09_{x}} $                 & $-0.580 $ & $-0.540 $ & $ $ \\
\cite{dietzel09} & $ $ & $ $                     & $ $ & $ $ & $-0.415$ \\
$ $ & $ $ & $ $ & $ $ & $ $ & $ $ \\
\hline
$ $ & $ $ & $ $ & $ $ & $ $ & $ $ \\
\cite{rana12} & ${\rm Mg-DOBDC}$ & ${\rm LDA}$ & $-0.542 $ & $-0.531 $ & $ $ \\
$ $ & $ $ & ${\rm PBE} $                       & $-0.228 $ & $-0.209 $ & $ $ \\
$ $ & $ $ & ${\rm DFT-D2} $                    & $-0.450 $ & $-0.411 $ & $ $ \\
$ $ & $ $ & ${\rm vdW-DF}     $               & $-0.528 $ & $-0.490 $ & $ $ \\
$ $ & $ $ & ${\rm vdW-DF2} $                  & $-0.500 $ & $-0.479 $ & $ $ \\
$ $ & $ $ & ${\rm vdW-optB86b} $               & $-0.582 $ & $-0.559 $ & $ $ \\
$ $ & $ $ & ${\rm vdW-optB88} $                & $-0.575 $ & $-0.547 $ & $ $ \\
$ $ & $ $ & ${\rm vdW-optPBE} $                & $-0.608 $ & $-0.593 $ & $ $ \\
\cite{valenzano11} & $ $ & ${\rm B3LYP+D}$     & $-0.430 $ & $-0.393 $ & $ $ \\
$ $ & $ $ & ${\rm MP2:B3LYP+D}$ & $-0.480 $ & $-0.443 $ & $ $ \\
\cite{rana12} & $ $ & $ $                      & $ $ & $ $ & $-0.458 \pm 0.048 $ \\
$ $ & $ $ & $ $ & $ $ & $ $ & $ $ \\
\hline
$ $ & $ $ & $ $ & $ $ & $ $ & $ $ \\
\cite{rana12} & ${\rm Ni-DOBDC}$ & ${\rm PBE}$ & $-0.124 $ & $-0.091 $ & $ $ \\
$ $ & $ $ & ${\rm DFT-D2} $                    & $-0.361 $ & $-0.317 $ & $ $ \\
$ $ & $ $ & ${\rm vdW-DF}     $               & $-0.428 $ & $-0.392 $ & $ $ \\
$ $ & $ $ & ${\rm vdW-DF2} $                  & $-0.405 $ & $-0.358 $ & $ $ \\
$ $ & $ $ & ${\rm vdW-optB86b} $               & $-0.504 $ & $-0.473 $ & $ $ \\
$ $ & $ $ & ${\rm vdW-optB88} $                & $-0.496 $ & $-0.447 $ & $ $ \\
$ $ & $ $ & ${\rm vdW-optPBE} $                & $-0.516 $ & $-0.483 $ & $ $ \\
\cite{valenzano11} & $ $ & ${\rm B3LYP+D}$     & $-0.403 $ & $-0.368 $ & $ $ \\
$ $ & $ $ & ${\rm MP2:B3LYP+D}$ & $-0.455 $ & $-0.420 $ & $ $ \\
\cite{rana12} & $ $ & $ $                      & $   $ & $ $ & $-0.410 \pm 0.016 $ \\
$ $ & $ $ & $ $ & $ $ & $ $ & $ $ \\
\hline
$ $ & $ $ & $ $ & $ $ & $ $ & $ $ \\
\cite{rana12} & ${\rm Co-DOBDC}$ & ${\rm LDA}$ & $-0.408 $ & $-0.379 $ & $ $ \\
$ $ & $ $ & ${\rm PBE} $                       & $-0.100 $ & $-0.086 $ & $ $ \\
$ $ & $ $ & ${\rm DFT-D2} $                    & $-0.322 $ & $-0.301 $ & $ $ \\
$ $ & $ $ & ${\rm vdW-DF}     $               & $-0.407 $ & $-0.386 $ & $ $ \\
$ $ & $ $ & ${\rm vdW-DF2} $                  & $-0.375 $ & $-0.337 $ & $ $ \\
$ $ & $ $ & ${\rm vdW-optB86b} $               & $-0.455 $ & $-0.419 $ & $ $ \\
$ $ & $ $ & ${\rm vdW-optB88} $                & $-0.452 $ & $-0.414 $ & $ $ \\
$ $ & $ $ & ${\rm vdW-optPBE} $                & $-0.477 $ & $-0.453 $ & $ $ \\
\cite{rana12} & $ $ & $ $                      & $ $ & $ $ & $-0.370 \pm 0.020 $ \\
$ $ & $ $ & $ $ & $ $ & $ $ & $ $ \\
\hline
$ $ & $ $ & $ $ & $ $ & $ $ & $ $ \\
\cite{rana12} & ${\rm Cu-HKUST}$ & ${\rm LDA}$ & $-0.326 $ & $-0.317 $ & $ $ \\
$ $ & $ $ & ${\rm PBE} $                       & $-0.092 $ & $-0.097 $ & $ $ \\
$ $ & $ $ & ${\rm DFT-D2} $                    & $-0.233 $ & $-0.192 $ & $ $ \\
$ $ & $ $ & ${\rm vdW-DF} $                   & $-0.283 $ & $-0.241 $ & $ $ \\
$ $ & $ $ & ${\rm vdW-DF2} $                  & $-0.264 $ & $-0.223 $ & $ $ \\
$ $ & $ $ & ${\rm vdW-optB86b} $               & $-0.305 $ & $-0.263 $ & $ $ \\
$ $ & $ $ & ${\rm vdW-optPBE} $                & $-0.329 $ & $-0.287 $ & $ $ \\
\cite{rana12} & $ $ & $ $                      & $ $ & $ $ & $-0.246 \pm 0.085 $ \\
$ $ & $ $ & $ $ & $ $ & $ $ & $ $ \\
\hline
\hline
\end{tabular}
\end{center}
\caption{CO$_{2}$-MOF binding energies, $E_{\rm bind}$, and heats of adsorption ($T = 300$~K), $\Delta H$, 
         calculated with different DFT exchange-correlation functionals. vdW-optB86b, vdW-optB88, and 
         vdW-optPBE represent variants of the original vdW-DF and vdW-DF2 nonlocal functionals~\cite{dion04,lee10b} 
         due to Klime\v{s} and co-workers~\cite{optex}. The vdW-C09$_{x}$ variant is due to Cooper~\cite{cooper10}, 
         and the vdW-PBE one is based on the PBE exchange functional~\cite{pbe} (see Sec.~\ref{subsec:vdw}). 
         Experimental values, $\Delta H^{\rm exp}$, correspond to measured heats of adsorption,  
         and all energies are given in units of eV per molecule.}
\end{table*}

\subsubsection{MOF}
\label{subsec:mof-ccs}
The interplay between open-metal site MOF and CO$_{2}$ molecules is dominated by dispersion
interactions, which represent about the $40-60$~\% of the total gas adsorption energy~\cite{ji14,valenzano10}.
Besides dispersion, electrostatic and orbital interactions are also important in shaping the
gas affinity of MOF. In particular, puntual charges localized in the unsaturated metal sites
polarize the CO$_{2}$ molecules inducing an electric dipole and a bend distorsion in 
them~\cite{ji14,poloni12,park12b,koh13}. When low-energy empty $d$-levels are present a forward
donation of electrons from CO$_{2}$ to the metal atoms occurs, which tends to increase the
electrostatic contribution to the gas binding~\cite{park12b,ji14,poloni14}. This last effect
is particularly important in MOF-74 and BTT species (see Sec.~\ref{subsec:mofs}) containing
Ti and V open-metal sites, where the $\sigma^{*}$ antibonding state that results from the
hybridization of oxygen CO$_{2}$ lone pairs and metal $d_{z}^{2}$ orbitals remains
unoccupied~\cite{poloni14}. In the case of heavily loaded MOF, quadrupolar CO$_{2}$-CO$_{2}$
interactions turn out to be also very important.

\begin{figure}[h]
\centerline{
\includegraphics[width=1.00\linewidth]{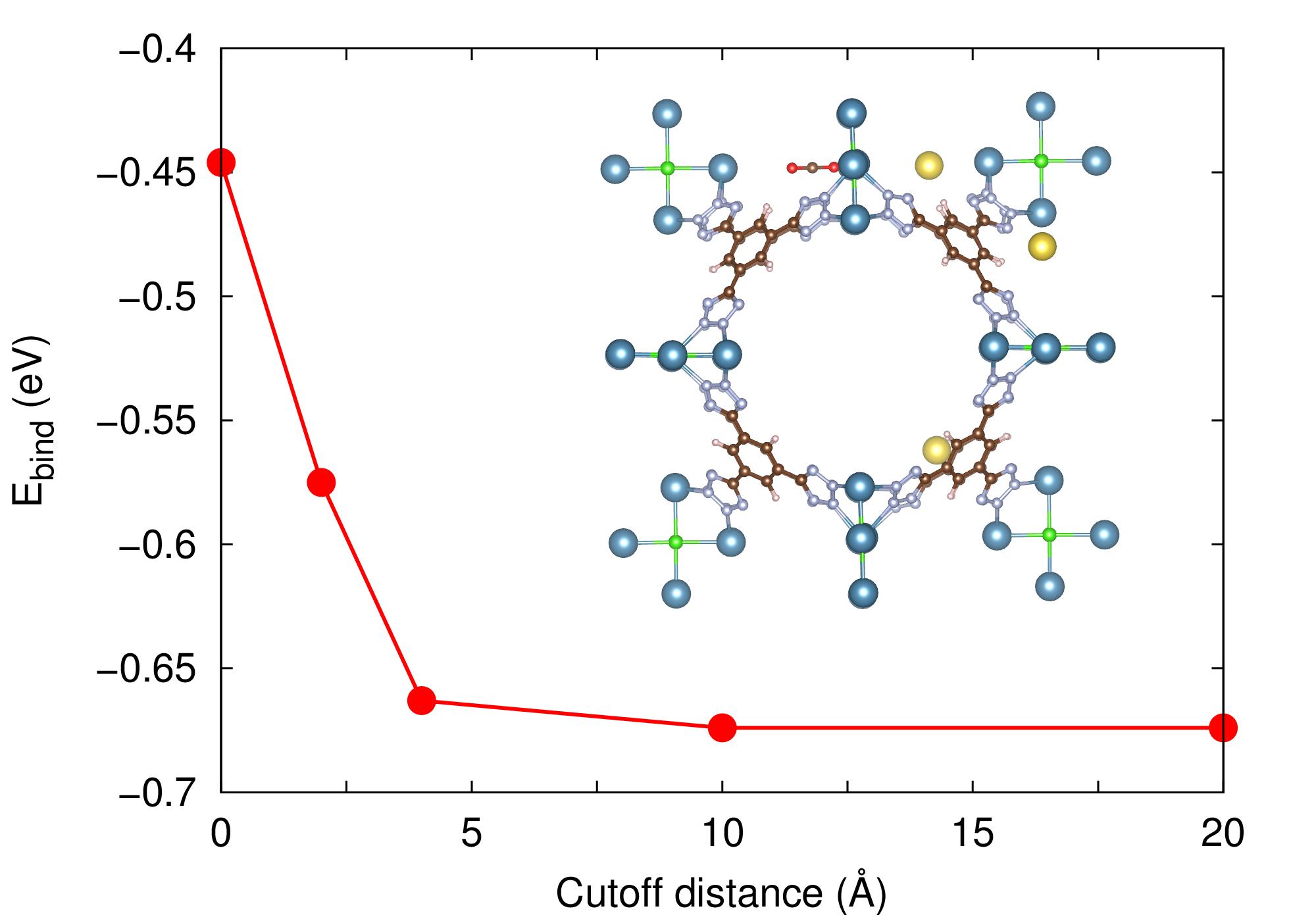}}
\caption{Binding energy of a CO$_{2}$ molecule on Ca-BTT computed with
the PBE+D2 approach, expressed as a function of the cutoff distance for the pairwise
corrections (adapted from Ref.~\cite{poloni12}). The CO$_{2}$-MOF system
in which calculations are performed is represented also in the figure.}
\label{mof-co2-dispersion}
\end{figure}

Next, we review recent DFT benchmarking works done on the screening of MOF for
applications in CO$_{2}$ capture and sequestration. Due to the crucial relevance of long-range
interactions in CO$_{2}$-MOF systems, we disregard here computational studies considering small organic
molecules as model GAM (see, for instance, articles~\cite{vogiatzis09,witte14} where the
CO$_{2}$-pyridine and CO$_{2}$-benzene interactions are investigated) since proper convergence
of long-range dispersion forces is achieved within distances of at least $\sim 10$~\AA~. This fact
is illustrated in Fig.~\ref{mof-co2-dispersion}, which has been adapted from work~\cite{poloni12},
where the DFT binding energy of a CO$_{2}$ molecule on a MOF containing Ca open metal sites is 
represented as a function of the cutoff distance that is employed in the calculation of the 
dispersion interactions. 

In Table~VI, we enclose the results of gas binding energy and enthalpy of adsorption, $\Delta H$,
calculations performed by Poloni \emph{et al.}~\cite{poloni12} and Rana \emph{et al.}~\cite{rana12}
in different MOF, using standard and dispersion corrected DFT functionals. For present benchmarking
purposes, comparisons with respect to quantum chemistry calculations performed in cluster-size model
systems may result not meaningful because of the convergence reasons explained above, and thus the 
DFT outcomes are compared with experimental heats of adsorption. Enthalpies of adsorption at ambient 
temperature can be estimated with the formula $\Delta H = E_{\rm bind} + \Delta E_{\rm ZPE} + \Delta E_{\rm TE}$. 
Here, $\Delta E_{\rm ZPE}$ and $\Delta E_{\rm TE}$ represent the zero-point energy (i.e., $E_{\rm ZPE} = 
\sum_{i} \hbar \omega_{i}/2$, $\lbrace \omega_{i} \rbrace$ being the corresponding vibrational phonon
frequencies) and the thermal energy (i.e., $E_{\rm TE} = E_{\rm vib} + E_{\rm rot} + E_{\rm transl} + k_{B}T$
for the CO$_{2}$ gas phase -where $k_{B}T$ accounts for the energy of an ideal gas-, 
and  $E_{\rm TE} = E_{\rm vib}$ for the framework with and without the adsorbate) differences, 
respectively, between the joint and disjoint CO$_{2}$-MOF systems. 

Upon comparison of the computed and measured $\Delta H$ values shown in Table~VI, we can draw the following 
two conclusions: first, standard DFT methods are far from reproducing the experimental heat of adsorption 
trends, with LDA~(PBE) presenting large underestimation~(overestimation) bias; and second, among the two 
families of considered dispersion corrected approaches, i.e., semiempirical Grimme's and vdW-$X$ (where 
$X$ indicates the choice of the $E_{x}^{\rm GGA}$ functional -see Sec.~\ref{subsec:vdw}-), the latter always
provides the better agreement with respect to experiments (in particular, the vdW-DF and vdW-DF2 variants 
show null $\Delta H - \Delta H^{\rm exp}$ discrepancies within the numerical uncertainties in most of the 
cases). It is worth noticing that the value of the $\Delta E_{\rm ZPE}$ and $\Delta E_{\rm TE}$ differences
generally amount to less than $\sim 50$~meV, hence the major contribution to $\Delta H$ comes from
the $E_{\rm bind}$ term. Also we note that the structural traits predicted in CO$_{2}$-MOF systems with 
vdW-$X$ methods, not shown here, present overall good agreement with the experiments.

In Table~VI, we also include the $\Delta H$ results obtained by Valenzano \emph{et
al.}~\cite{valenzano11} with an hybrid MP2:B3LYP-D approach~\cite{svelle09,tosoni10}.
In that scheme, MP2 calculations are carried out first in a cluster model system representing the 
adsorption site, and a long-range correction is added afterwards. The long-range correction is 
defined as the B3LYP-D energy difference between the extended and model cluster systems. 
(The MP2:B3LYP-D method can be likewise understood as starting from a B3LYP-D calculation
for the full periodic structure and adding a high-level correction for the adsorption site.)
Valenzano's MP2:B3LYP-D results obtained in the Mg- and Ni-DOBDC systems actually are in the same
excellent agreement with experimental data than found with the vdW-DF and vdW-DF2 methods
(i.e., null $\Delta H - \Delta H^{\rm exp}$ discrepancies within the numerical uncertainties).
Also, they represent an improvement with respect to primary B3LYP-D calculations~\cite{valenzano11}.
On one hand, Valenzano's MP2:B3LYP-D work comes to reinforce the accuracy of vdW-DF methods 
in describing the gas uptake properties of MOF. On the other hand, it demonstrates the reliability 
and efficiency of the intuitive MP2:B3LYP-D approach for undertaking benchmark and GAM design studies.

\emph{Conclusions.-}~In light of the compiled results and analysis, it can be concluded 
that the use of standard DFT functionals must be avoided in the study of the CO$_{2}$ adsorption 
features of MOF. Meanwhile, the suite of vdW-DF methods currently represent the best choice
for carrying out computational studies of such a type, both in terms of computational expense
and reliability. In this regard, we would like to mention that in order to complete our knowledge of 
the general performance of DFT methods in the simulation of CO$_{2}$-MOF systems it will be highly desirable 
to consider meta-GGA functionals (e.g., M05-2X and M06-2X) in prospective studies. Also, it would be 
very interesting to perform advanced RPA-DFT and QMC benchmark calculations in those same systems for 
complementing the comparisons reported against experiments, and determining more precisely the role 
of the different energy contributions to the binding of molecules (e.g., many-body dispersion effects). 
Meanwhile, merging quantum chemistry methods (e.g., MP2) with periodic DFT simulations that incorporate 
long-range dispersion interactions, appears to be an effective strategy for calculating accurate binding 
energies in situations where the selected cluster and extended systems are physically alike.

\section{Discussion and Prospective Work}
\label{sec:discussion}
\begin{figure}[h]
\centerline{
\includegraphics[width=1.00\linewidth]{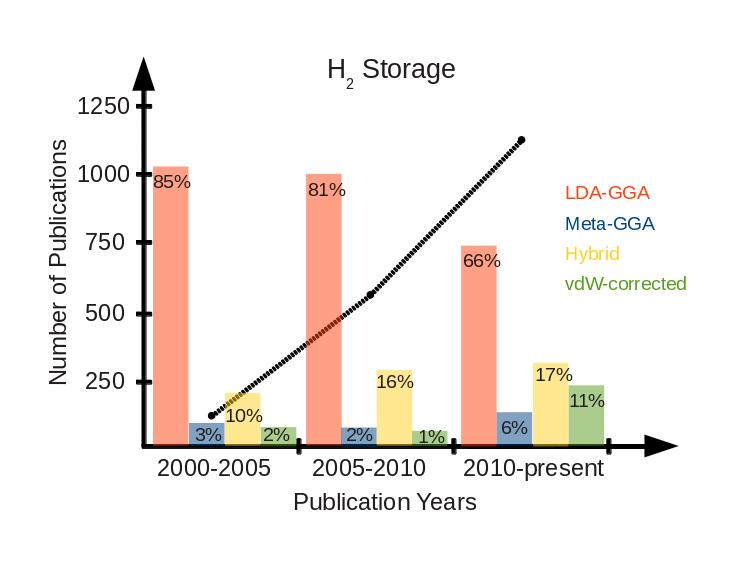}}
\vspace{-0.50cm}
\caption{Number of DFT-based ``H$_{2}$ storage'' articles
         published from 2000 onwards, classified according
         to the employed $E_{xc}$ approximation (our estimation
         relies on data extracted from the ``Web of Knowledge'', 
         December 2014). The solid dots represent the total number
         of works published in different years intervals, and the
         boxes the percentage corresponding to each DFT variant.}
\label{discussion-H2-storage}
\end{figure}

\begin{figure}[h]
\centerline{
\includegraphics[width=1.00\linewidth]{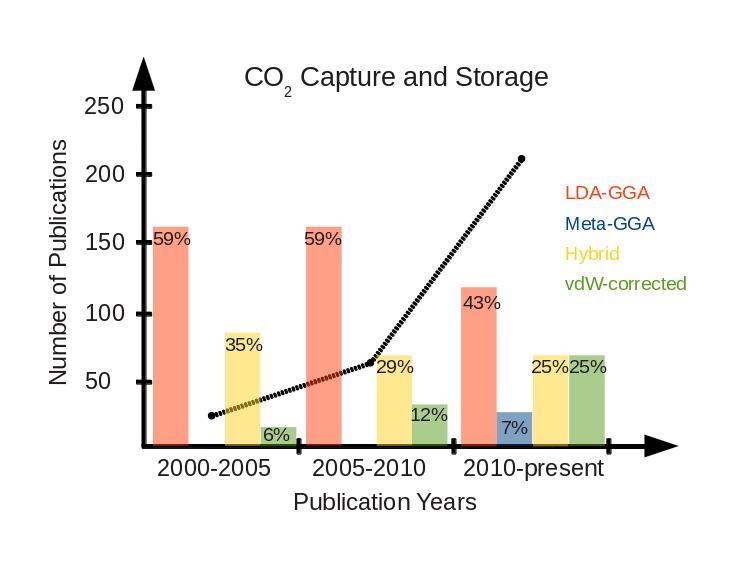}}
\vspace{-0.50cm}
\caption{Number of DFT-based ``Carbon capture and storage'' articles
         published from 2000 onwards, classified according
         to the employed $E_{xc}$ approximation (our estimation
         relies on data extracted from the ``Web of Knowledge'', 
         December 2014). The solid dots represent the total number
         of works published in different years intervals, and the
         boxes the percentage corresponding to each DFT variant.}
\label{discussion-CCS}
\end{figure}

From all the results and explanations presented in Sec.~\ref{sec:assessmentDFT} it can be concluded 
that standard DFT methods (i.e., LDA and GGA) generally provide results on the fixation 
of H$_{2}$ and CO$_{2}$ molecules in CN and MOF that are correct only at the qualitative level. Actually,
in the sole analyzed case of hydrogen binding to chemically functionalized MOF coherent agreement between DFT-GGA
results and highly accurate quantum chemistry calculations has been reported by several authors (see
Sec.~\ref{subsec:mof-h2}). However, in that particular case full consistency between DFT results obtained 
in cluster size and periodic H$_{2}$-MOF systems is still lacking, and further investigations are required 
for determining the precise role of dispersion interactions. Therefore, our general recommendation 
is to avoid using standard DFT methods in first-principles modeling of H$_{2}$-storage and CCS materials 
when pursuing accurate binding energies and geometries. In stark contrast to this our advice, most of the 
computational studies published to date on gas-adsorption topics have relied heavily on the outcomes of 
DFT-LDA and DFT-GGA simulations (i.e., about the $80-60$~\% of them in the last $14$ years).
In Figs.~\ref{discussion-H2-storage} and~\ref{discussion-CCS}, we show an estimation of 
the total number of DFT works reported in the areas of H$_{2}$ storage and CCS GAM modeling, 
classified according to the employed DFT functional and publication date. 

In the case of hydrogen storage (see Fig.~\ref{discussion-H2-storage}), we appreciate an 
important recent decline in the relative number of modeling studies performed with LDA and GGA functionals, 
as compared to the trend observed during the first decade of the present century (i.e., a 
decrease of the $\sim 20$~\%). On the other side, the relative number of DFT works based on 
dispersion corrected schemes has increased significantly over the same period of time (i.e., 
from $\sim 1$~\% to $\sim 10$~\%). This last datum reflects, on one hand, the improved 
feasibility of electronic band structure methods incorporating van der Waals interactions 
(and the increasing availability through commercial packages) and, on the other, the awareness of the 
importance of this type of interactions within the community of materials scientists. Also, we notice 
that the popularity of hybrid and meta-GGA functionals have grown considerably within the last 
ten years. Nevertheless, all these trends refer to percentages and the truth is that, since the 
total number of published articles has grown almost linearly since year $2000$, the total number 
of computational works that are prone to revision (i.e., those performed with standard DFT 
methods) has actually increased during the last years. 

As for the modeling of CCS GAM (see Fig.~\ref{discussion-CCS}), we also acknowledge a steady 
increase in the total number of published works and a significant recession in the percentage 
of recent LDA and GGA studies (i.e., a decrease from $\sim 60$~\% to $\sim 40$~\% in the last 
four years). Meanwhile, the number of dispersion corrected DFT works has increased remarkably 
in the last years, reaching a maximum peak of $25$~\% recently, while the use of hybrid 
functionals has been maintained more or less constant around $30$~\%. Also, meta-GGA functionals 
are becoming increasingly more popular although these are still the least preferred among all the
considered DFT variants.
 
The ultimate DFT tendencies revealed in both H$_{2}$ storage and CCS GAM modeling fields are 
quite similar, namely there is a firm surge in the use of hybrid and dispersion corrected approaches in 
detriment to those of LDA and GGA. This fact shows that the outcomes of complex DFT benchmark 
studies are being assimilated progressively by the community of materials scientists specialized in 
GAM applications. In spite of this positive conclusion, the total number of recent GAM design works that 
still rely exclusively on LDA and GGA calculations is, in our opinion, unjustifiably too large.     

In the Introduction section of this review we commented on the two main threats that customary 
DFT approaches pose to the design of GAM, namely the difficulties in (i)~accounting simultaneously 
for the long range electron-electron exchange and correlations, both of which are omnipresent 
in gas-adsorption phenomena, and (ii)~reproducing many-body energy effects and Coulomb screening  
in extended systems. 

Regarding issue (i), we have demonstrated in Sec.~\ref{sec:assessmentDFT} 
that good agreement with respect to quantum chemistry calculations and experimental data is  
achieved when well-balanced exchange-correlation functionals are employed that include van der 
Waals interactions, on one hand, and correct to some extent for the inherent electron 
self-interaction errors, on the other. This is the case, for instance, of the mixed BLYP-D3 method 
which in the CO$_{2}$/2-methylpyridine system has been shown to perform at the same level of accuracy 
than CCSD(T)/CBS (see Sec.~\ref{subsec:carbon-gam-ccs} and work~\cite{mo13}). Also, the so-called
$\omega$B97X-D approach, based on a range-separated hybrid and dispersion corrected DFT functional 
(see Sec.~\ref{subsec:hybrid}), has been shown to reproduce correctly the binding energy trends measured 
for H$_{2}$ molecules on metal-BTT complexes (see Sec.~\ref{subsec:mof-h2} and work~\cite{sumida13}).
We consequently argue that the \emph{safest} options for undertaking first-principles computational
work on gas-adsorption applications and the design of GAM are to use DFT functionals which incorporate, 
either in an exact or approximate way, long range electron-electron exchange and correlations. 
The performance of these ``full long range corrected'' (FLRC) DFT functionals has been tested 
very recently by several authors in standard benchmark data sets, and very promising results 
have been attained in all the cases (e.g., see studies~\cite{gordon08,lin13,mardirossian14,vydrov12}). 

An interesting aspect of FLRC functionals is that these can be constructed in principle by taking any 
pure or long range corrected hybrid DFT functional as a start, and subsequently adding the missing nonlocality 
of the correlation energy in the form of additional $E_{xc}$ terms. Also, the degree of sophistication and 
computational expense associated to FLRC functionals can be chosen almost at wish by adopting simpler 
or more complex exchange-correlation correction schemes. For instance, a computationally inexpensive 
FLRC solution may consist in combining the semiempirical dispersion correction approach due to Grimme with 
any hybrid or meta DFT functional of one's personal taste (e.g., the already employed $\omega$B97X-D, 
TPSS-D3, M06-D3, and BLYP-D3 functionals). A superior FLRC blend, both in terms of accuracy and 
computational expense, could be achieved by considering full nonlocal functionals like vdW-DF or VV10 
rather than semiempirical correlation correction schemes (see for instance work~\cite{vydrov12}). 
In conclusion, we strongly recommend to use FLRC functionals in future \emph{ab initio} modeling of GAM 
and to analyze their performance comprehensively in prospective DFT benchmark studies. 

Regarding the second customary DFT challenge~(ii) mentioned above and explained in Sec.~\ref{subsec:dftchallenges}, 
many-body energy and Coulomb screening effects can vary considerably the polarizabilities, and consequently 
the forces, of gas molecules interacting with extended surfaces and other molecules~\cite{ruiz12,schimka10}.     
The RPA-DFT and DFT+MBD methods emerge as the two DFT variants which can deal efficiently with these types 
of many-body effects (see Sec.~\ref{subsec:rpa}). In particular, the DFT+MBD method is very well suited 
for studying large periodic systems containing up to few hundreds of atoms~\cite{tkatchenko12b,ambrosetti14,distasio14}.   
Nevertheless, to the best of our knowledge, neither the RPA-DFT nor the DFT+MBD method has been applied yet 
to the simulation of hydrogen storage or carbon capture and sequestration problems, as shown by 
the lack of such studies in the comprehensive analysis presented in Sec.~\ref{sec:assessmentDFT}. 
Prospective RPA-DFT and DFT+MBD works targeting those GAM design areas are highly desirable for rationalizing 
further the causes underlying the discrepancies found between DFT-based and quantum chemistry methods [e.g., 
DMC and CCSD(T)], and in general for substantiating the role of many-body effects in the adsorption of gas molecules 
on atomically sparse environments. In fact, many-body DFT methods themselves present also a number of remaining 
challenges like it can be the derivation of an analytic expression for the calculation of atomic forces. 
(This ingredient is highly sought after for the realization of efficient geometry optimizations and molecular 
dynamics simulations.) Work on this and other directions have been already initiated~\cite{ambrosetti14,distasio14}, 
thereby we expect that the application of many-body DFT methods will become routinary in the near future. 
We animate computational scientists to start considering these advanced computational techniques in 
their planned studies of GAM.  

Concerning the generalization of DFT benchmark conclusions attained in cluster systems to realistic materials, 
we have presented evidence for a number of related issues like are the omission of long range 
dispersion interactions and the disguise of electronic band structure effects (see, Secs.~\ref{subsec:benchmarking} 
and~\ref{sec:assessmentDFT}). An effective way of getting rid of those potential drawbacks is to recover the 
translational invariance in the simulations and to perform all the required calculations in a same periodic 
system. This possibility naturally points to the suite of \emph{exact} ground-state quantum Monte Carlo methods 
(e.g., diffusion Monte Carlo -DMC-) as one of the most promising approaches for undertaking benchmark studies on 
the performance of DFT. In fact, the formalism of Bloch functions is already firmly established within 
QMC~\cite{rajagopal95,foulkes01} and different implementations of this approach are available 
in several open-source packages (e.g., CASINO~\cite{needs10}, QWALK~\cite{wagner09}, and QMCPACK~\cite{qmcpack}). Also, the 
computational cost of QMC calculations is orders of magnitude more favorable than those of other popular quantum chemistry 
methods. For these reasons, we envisage that QMC methods will play an increasingly more relevant role in prospective DFT 
benchmark studies of GAM. (Actually, the DMC method has already been applied to the study of metal hydrides~\cite{wu09,binnie10,pozzo08}
and metal oxides~\cite{wagner07,alfe06,binnie09}, two important families of materials within the context of hydrogen storage 
and carbon capture.) It is worth noticing, however, that QMC methods are neither exempt of some important technical 
problems. These are essentially related to the tediousness found in the computation of atomic forces~\cite{wagner10,badinski10} 
and the correction of numerical bias introduced by the nodal surface approximation~\cite{foulkes01}. 

An alternative to using periodic quantum chemistry methods, which is computationally more feasible but also 
tentative, may consist in designing model cluster systems which mimic the electronic structure of targeted 
extended GAM (see Sec.~\ref{subsec:carbon-gam-ccs}) and where all the hierarchical calculations are subsequently 
performed. In following this approach, one should correct somehow for the electronic long-range exchange 
and correlations in the final adsorption energies. This could be done \emph{ad hoc} by carrying out additional 
calculations in both periodic and cluster systems with a suitable first-principles method (e.g., FLRC functionals 
like $\omega$B97X-D and BLYP-D3). Such a proposed long-range correction scheme is very much on the spirit 
of hybrid DFT:QCM approaches, which have been demonstrated to be successful on the simulation of extended systems 
(e.g., see Sec.~\ref{subsec:mof-ccs} and works~\cite{svelle09,tosoni10}). The reliability of this alternative 
benchmark approach, however, has not been yet fully assessed hence further work on this direction is certainly needed 
before claiming any progress.

\section{General Conclusions}
\label{sec:conclusions}
In this critical review, we have presented abundant evidence showing that both electronic long-range 
exchange and long-range correlations play a decisive role in the adsorption of H$_{2}$ and 
CO$_{2}$ molecules on carbon-based materials and MOF. Even in situations where the calculated 
gas binding energies turn out to be large (i.e., $E_{\rm bind} \ge 0.1$~eV), the same conclusion 
holds to be true. Consequently, DFT exchange-correlation functionals employed in the modeling of GAM 
must incorporate features which somehow correct for standard self-interaction errors and simultaneously 
reproduce dispersion forces. DFT $E_{xc}$ functionals of this class include the brand-new
BLYP-D3, $\omega$B97X-D, and vdW-optB88 variants, to cite a few. Other similar ``full long range 
corrected'' functionals can be tailored in principle by combining hybrid and dispersion correction 
schemes of one's personal taste, producing so a suite of \emph{safe} DFT approaches which can range 
widely on versatility and computational expense.  

As for standard DFT functionals (i.e., LDA and GGA), since these completely disregard electronic
long-range exchange and correlations, we recommend not to use them in the modeling of GAM 
and simulation of gas adsorption phenomena in general. In some particular circumstances LDA and GGA 
approaches may provide the correct qualitative answers, however this is likely to occur fortuitously 
as a result of large error cancellations. Regrettably, current trends realized in modeling of GAM 
show that the use of LDA and GGA functionals is still quite widespread. We expect to motivate a change 
on this tendency with the present critical review. Also, we appeal to bring into new examination 
reported standard DFT predictions on GAM which are in conflict with observations. 

Concerning prospective DFT work, we animate researchers to consider the RPA-DFT and DFT+MBD 
methods in future studies of H$_{2}$ storage and carbon capture materials. The reason for this is to 
substantiate with precision the role of many-body energy and Coulomb screening effects on the 
estimation of gas binding energies and equilibrium geometries. This is a completely new direction to take 
in modeling of GAM and, despite that the applicability of many-body DFT-based approaches remains yet limited, 
we believe that it will help in understanding further the origins of the discrepancies found between DFT 
and quantum chemistry results.      

Finally, we recommend to be cautious in generalizing the conclusions of benchmark studies 
performed in model cluster systems to realistic GAM. In fact, the density of electronic 
states around the Fermi level and the HOMO and LUMO orbitals in chemically similar 
cluster and extended systems, even when considering medium size model complexes, can be very 
different. Those circumstances easily can translate into completely different gas-GAM governing 
interactions. Also, electronic long-range interactions present in extended materials generally
turn out to be disguised in model cluster systems. The safest strategy for avoiding these potential 
scaling problems is to consider quantum chemistry methods which, alike to DFT, can handle the simulation 
of periodic systems. In this context, quantum Monte Carlo emerge as one the most effective suite 
of \emph{exact} ground-state methods for computation of gold-standard benchmarks in GAM.

\acknowledgments
This research was supported under the Australian Research Council's Future Fellowship funding scheme 
(project number RG134363 and RG151175). The author would like to thank S. A. Shevlin for useful 
discussions on the organization of this review and for help in preparing Fig.~\ref{figintro-1}.

\end{document}